\documentclass{appolb}
\usepackage{graphicx}

\newcommand{\bp}{\mbox{\boldmath $p$}}
\newcommand{\bq}{\mbox{\boldmath $q$}}

\newcommand{\bP}{\mbox{\boldmath $P$}}

\newcommand{\bk}{\mbox{\boldmath $k$}}

\newcommand{\ket}[1]{ {#1} \rangle}

\renewcommand{\thetable}{\arabic{table}}

\begin{document}
\title{From transition electromagnetic form factors $\gamma^* \gamma^* \eta_c(1S,2S)$ to the production of $\eta_c(1S,2S)$ at the LHC%
\thanks{Presented at XXVI Cracow EPIPHANY Conference, LHC Physics: Standard Model and Beyond, 7-10 January 2020}%
}
\author{Antoni Szczurek
\address{Institute of Nuclear Physics Polish Academy of Sciences,\\
ul. Radzikowskiego 152, PL-31342 Krak\'ow, Poland,\\
Faculty of Mathematics and Natural Sciences,University of Rzesz\'ow,\\
ul. Pigonia 1, PL-35-310 Rzesz\'ow, Poland}
}
\maketitle
\begin{abstract}
We review our recent results for production of $\eta_c(1S)$ and 
$\eta_c(2S)$ in the $\gamma^* \gamma^* \to \eta_c(1S,2S)$ fusion and 
in proton-proton collisions via gluon-gluon fusion.
The quarkonium wave functions are calculated by solving
Schr\"odinger equation for different $c \bar c$ potentials.
Using Terentev prescription the light-cone wave functions are obtained.
The light-cone wave functions are used then to calculate
$\gamma^* \gamma^* \to \eta_c$ transition form factors.
The theoretical results are compared to the Belle experimental data
for $\eta_c(1S)$. In addition we discuss our results for two-photon 
decay width.
We present also results of our calculations for proton-proton collisions
obtained within $k_T$-factorization approach for different unintegrated
gluon distributions. The results for hadroproduction of $\eta_c(1S)$ 
are compared to the LHCb experimental data.
\end{abstract}


\section{Introduction}

There has been a lot of interest recently in the exclusive production
of mesons via photon-photon fusion processes studied mainly at 
the $e^+ e^-$ colliders.
Such studies were motivated by the expectation that at 
large photon virtualities the measurements of the cross sections 
provide strong constrains in the probability amplitude for finding 
partons in the mesons 
\cite{Radyushkin:1977gp,Lepage:1979zb,Chernyak:1983ej}.
The meson - photon transition form factors are also of interest because 
of the role they play in the hadronic light-by-light contribution 
to the muon anomalous magnetic moment \cite{Jegerlehner:2017gek}.

A lot of attention has been paid to the case of pseudoscalar light meson
motivated by the experimental data from the CLEO, BaBar, Belle
and L3 Collaborations  for the $\pi^0$, $\eta$ and $\eta'$ production 
in $e^+ e^-$ collisions. These collaborations extracted the
transition form factor from single - tag events where only one of 
the leptons in the final state is measured. In this case, one of 
the photons is far off the mass shell, while the other is almost real. 
Such data allow to test the collinear factorization approach and 
the onset of the asymptotic regime, as well motivated the improvement 
of the theoretical approaches. 

Similar results have been obtained for the $\eta_c$ production. 
In this case, the $\eta_c$ mass provides a hard scale that justifies 
to use a perturbative approach even for zero virtualities. 
In the past this transition form factor 
was studied in different approaches, (although often
only for one virtual photon), such as: 
perturbative QCD \cite{Feldmann:1997te,Cao:1997hw}, 
lattice QCD \cite{DE2006,CLQCD}, 
non-relativistic QCD \cite{FJS2015,WWSB2018}, 
QCD sum rules \cite{LM2012}, as well
as from Dyson-Schwinger and Bethe-Salpeter equations \cite{CDCL2017}. 
In the light-front quark model (LFQM) the case
of one virtual and one real photon has been studied 
in \cite{GL2013,Ryu:2018egt}. 

\begin{figure}[h]
\centering\includegraphics[width=0.4\linewidth]{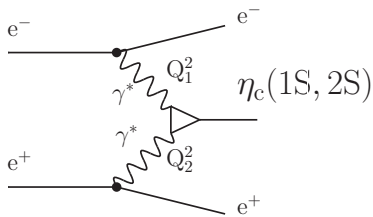}
\caption{Basic diagram for the $\gamma^* \gamma^* \to \chi_c$ coupling.}
\label{fig:diagram_gammagamma_etac}
\end{figure}

The quarkonia production reactions in hadronic collisions 
is also very interesting (see e.g. \cite{Lansberg:2019adr}). 
In Ref.\cite{BPSS2020}, we concentrated on the direct
hadroproduction of the ground state of the charmonium family,
$\eta_c(1S)$, and its first excited state $\eta_c(2S)$. 
Both are pseudoscalar particles of even charge parity $J^{PC} = 0^{-+}$. 
Like other $C$-even quarkonia, the dominant production mechanism is
through the $gg \to {\cal{Q}}$ gluon fusion $2 \to 1$ process.
For comparison in the standard collinear-factorization approach one must go 
to next-to-leading order (NLO) approximation to calculate 
the transverse momentum distribution of the quarkonium state 
and include $2 \to 2$ processes like $g g \to {\cal{Q}} g$. 
In the $k_T$-factorization approach \cite{UGD_GLR,UGD_CCH,UGD_CE}, 
the transverse momentum of the quarkonium originates from 
the transverse momenta of incident virtual gluons entering 
the hard $g^* g^* \to {\cal{Q}}$ process.

The $k_T$-factorization approach is especially appropriate in the
high-energy kinematics, where partons carry small momentum fractions 
of the incoming protons, mainly discussed in the framework of 
the BFKL formalism \cite{BFKL}. In our recent calculations we adopted 
\cite{BPSS2020} the color-singlet model, which treats the quarkonium as 
a two-body bound state of a heavy quark and antiquark. Such a formalism 
was used previously for the production of $\chi_{cJ}$ ($J=0,1,2$)
quarkonia (see e.g. Ref.~\cite{CS2018}), and a relatively good agreement
with data was obtained from an unintegrated gluon distribution (UGD),
which effectively includes the higher-order contributions.

\section{$\gamma^* \gamma^* \to \eta_c$ coupling}

\subsection{Nonrelativistic quarkonium wave functions}

\begin{figure}[h]
\centering\includegraphics[width=0.45\linewidth]{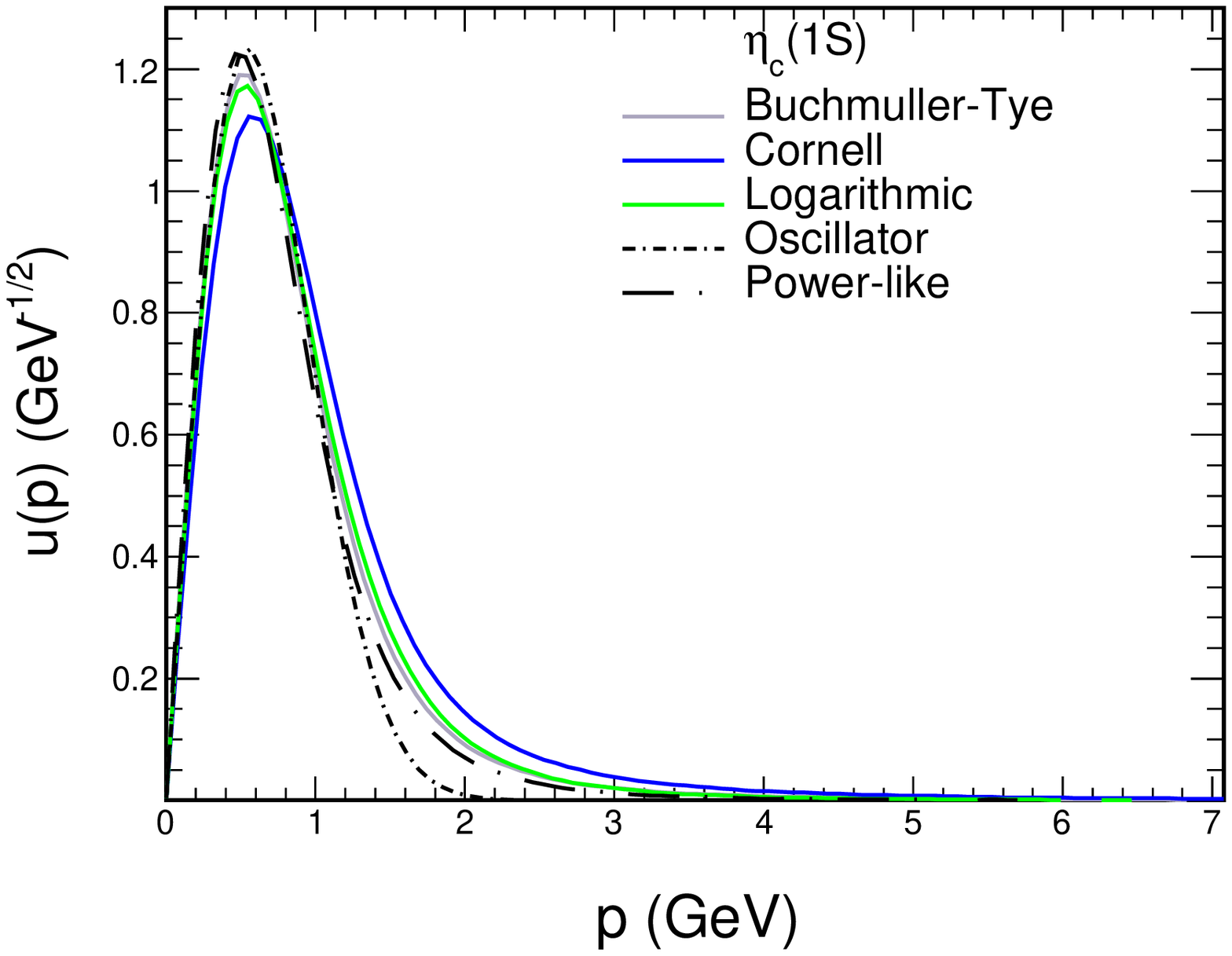}
\centering\includegraphics[width=0.45\linewidth]{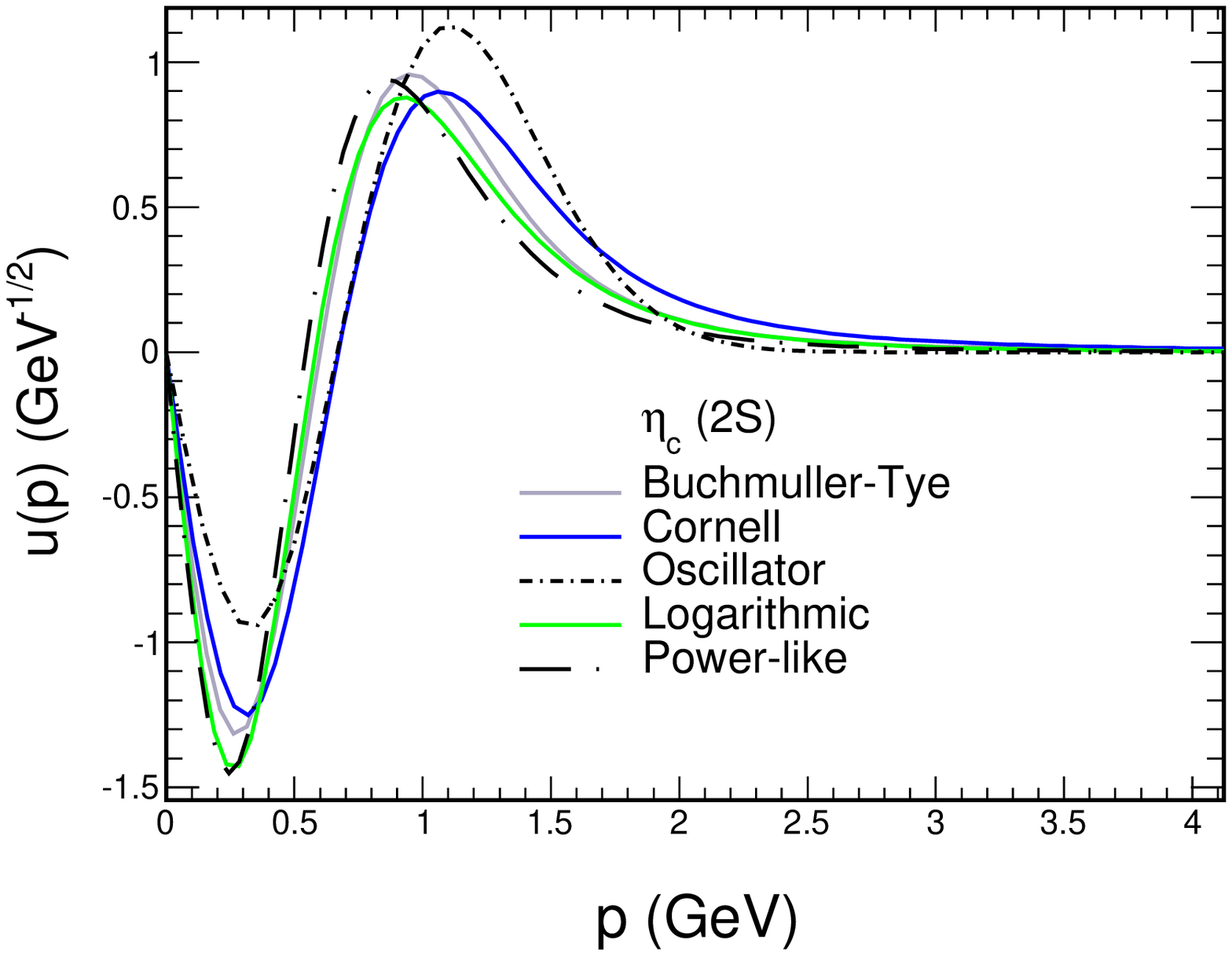}\\
\caption{ Radial momentum-space wave function for different potentials.
}
\label{fig:nrwf_momentum_space}
\end{figure}

The radial spatial wave functions were obtained by solving 
the Schr\"odinger equation \cite{CNKP2019}. Different potential models
known from the literature were used. The momentum wave functions can be
obtained then by calculating Fourier transform from the spatial
wave functions. In Fig.\ref{fig:nrwf_momentum_space} we show the 
resulting wave functions. One can observe some dependence on the
potential used in the calculation.

In our approach we treat the $\eta_c$ meson as a bound state of a charm quark 
and antiquark, assuming that the dominant contribution comes from 
the $c \bar c$ component in the Fock-state expansion:
\begin{eqnarray}
|\ket{\eta_c; P_+, \bP} = \sum_{i,j,\lambda, \bar \lambda}
{\delta^i_j \over \sqrt{N_c}} \, 
\int {dz d^2\bk \over z(1-z) 16 \pi^3} \Psi_{\lambda \bar \lambda}(z,\bk)
|\ket{c_{i \lambda}(z P_+ ,\bp_c)
\bar c^j_{\bar \lambda}((1-z)P_+,\bp_{\bar c})} + \dots
\nonumber \\
\end{eqnarray}
Here the $c$-quark and $\bar c$-antiquark carry a fraction $z$ and $1-z$ 
respectively of the $\eta_c$'s plus-momentum. The light-front helicites 
of quark and antiquark are denoted by $\lambda, \bar \lambda$, and
take values $\pm 1$.
The transverse momenta of quark and antiquark are
\begin{eqnarray}
\bp_c = \bk + z \bP \, , \quad \bp_{\bar c} = -\bk + (1-z) \bP \, .
\end{eqnarray}
The light-cone representation is obtained by 
Terentev's prescription \cite{Terentev:1976jk} valid for weakly 
bound systems.

\begin{figure}[h]
\centering\includegraphics[width=0.39\linewidth]{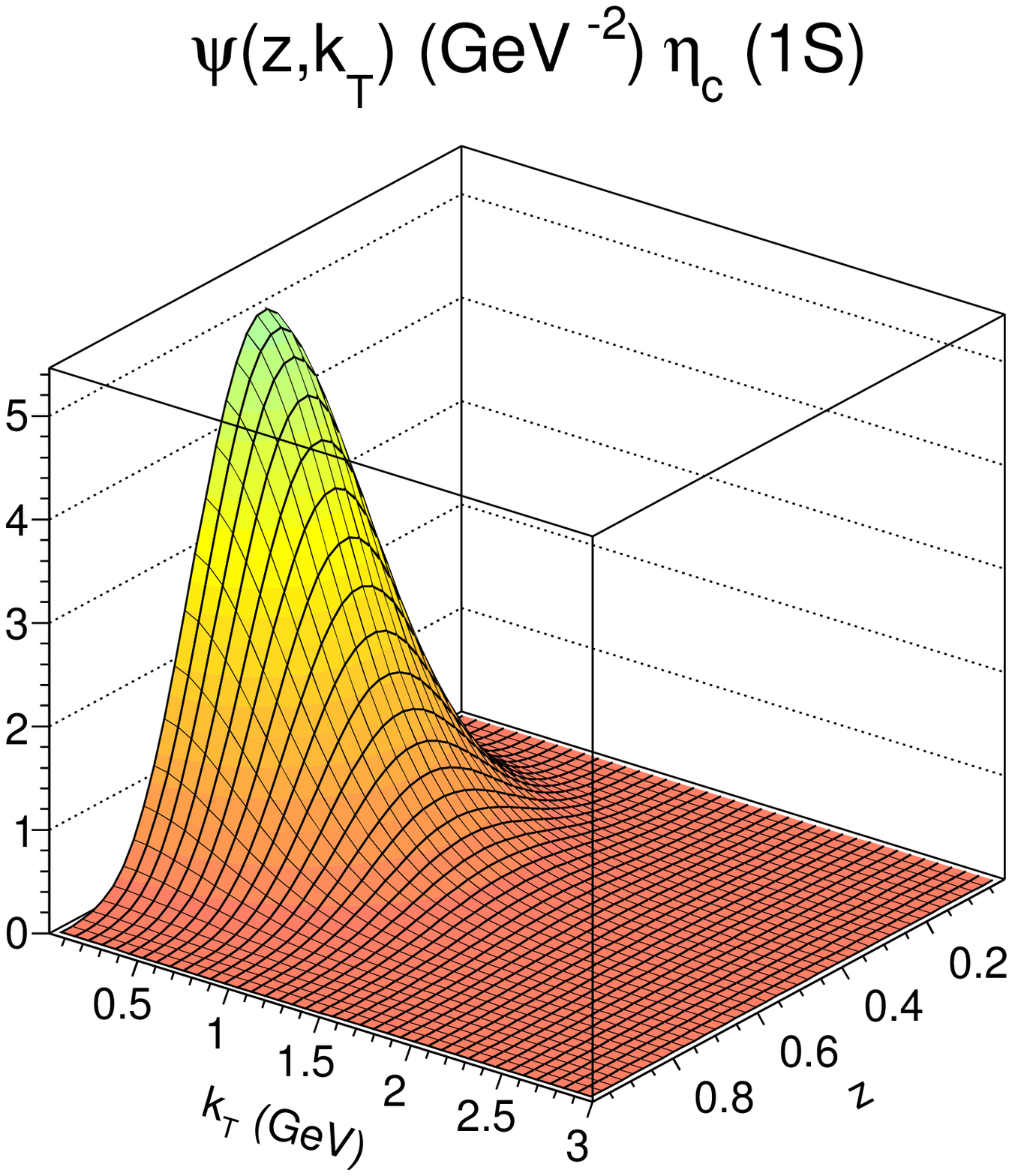}
\centering\includegraphics[width=0.39\linewidth]{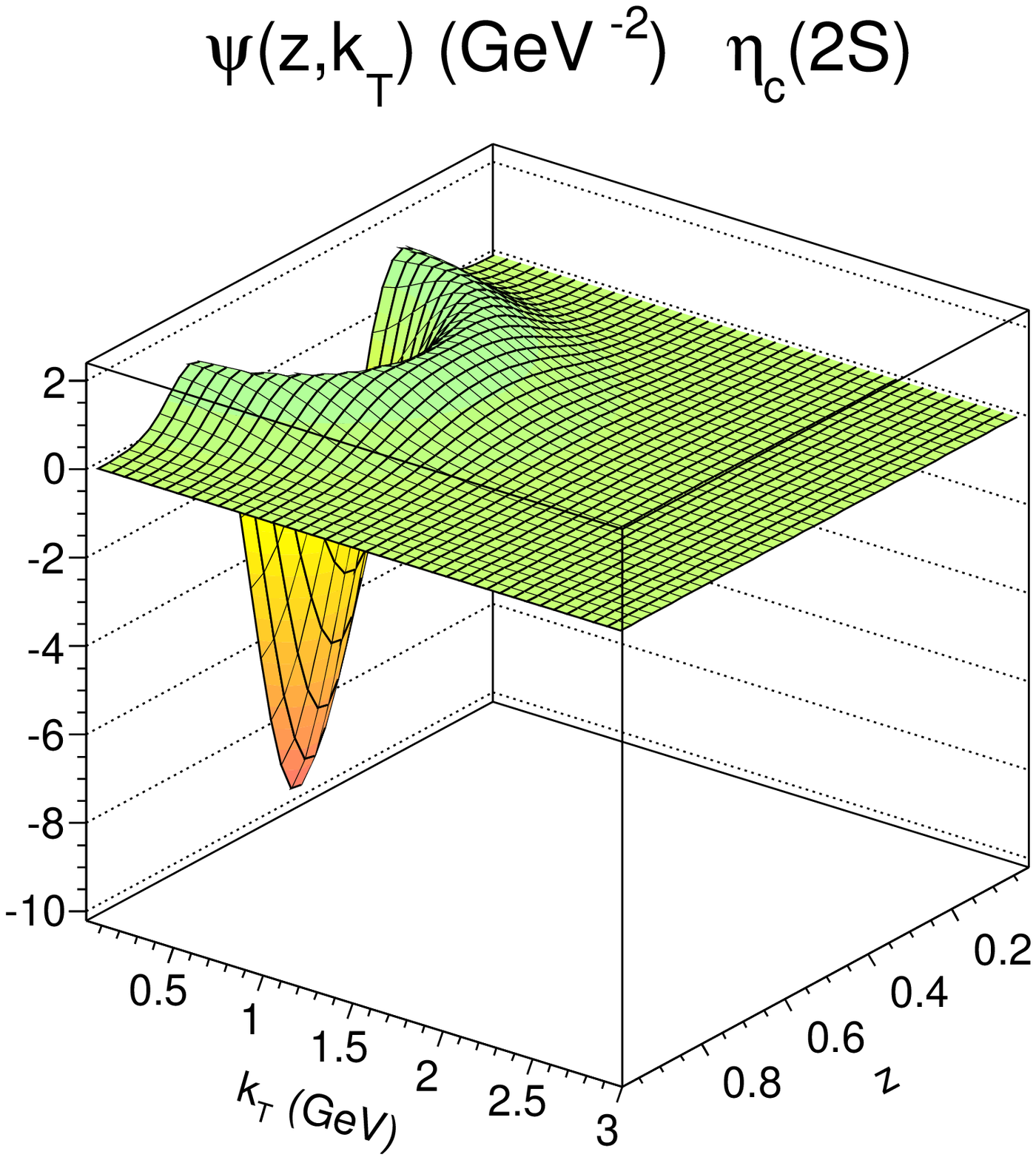}\\
\caption{ Radial light-front wave function for 
Buchm\"uller-Tye potential.}
\label{fig:LCWF}
\end{figure}

The resulting light-cone wave functions are shown in Fig.\ref{fig:LCWF}
for a selected $c \bar c$ potential specified in the figure caption.
%
%
According to the Terentev prescription \cite{Terentev:1976jk}:  
$\Rightarrow \textbf{p} = \textbf{k}, \quad p_z = (z - {1 \over 2}) M_{c \bar c}$,
\begin{eqnarray}
\psi(z,\textbf{k}) =
{\pi \over \sqrt{2 M_{c \bar c}}} { u(p) \over p} \, . \quad \nonumber
\end{eqnarray}

The value of the transition form factor at $Q_1^2, Q_2^2$ = 0 can be
calculated as:
\begin{eqnarray}
F(0,0) = e_c^2 \sqrt{N_c} \, 4m_c \cdot \int{dz d^2\textbf{k} \over z(1-z) 16 \pi^3}  {\psi(z,\textbf{k}) \over \textbf{k}^2 + m_c^2} \, . \nonumber
\end{eqnarray}
$F(0,0)$ is related to the two-photon decay width:
\begin{eqnarray}
\Gamma(\eta_c \to \gamma \gamma) = {\pi \over 4} \alpha^2_{\rm em} M_{\eta_c}^3 \, |F(0,0)|^2 \, . \nonumber
\end{eqnarray}

$F(0,0)$ can be rewriten in the terms of radial momentum space wave
function $u(p)$:
\begin{eqnarray}
F(0,0) = e_c^2 \sqrt{2 N_c} \, {2m_c \over \pi} \, \int_0^\infty {dp \, p  \, u(p) \over 
\sqrt{M_{c \bar c}^3} (p^2 + m_c^2)} \, {1 \over 2 \beta} \log\left({1+\beta \over 1 - \beta}\right)\, , \nonumber
\end{eqnarray}
In the non-relativistic (NR) limit, where $p^2/m_c^2 \ll 1, \beta \ll 1$,
and $2 m_c = M_{c \bar c} = M_{\eta_c}$, we obtain
\begin{eqnarray}
F(0,0) = e_c^2 \sqrt{N_c} \sqrt{2} {4 \over \pi \sqrt{M_{\eta_c}^5}} \int_0^\infty dp \, p \,  u(p) 
= e_c^2 \sqrt{N_c} {4 \,  R(0) \over \sqrt{\pi M_{\eta_c}^5}}\,, \nonumber
\end{eqnarray}
where $\beta = {p \over \sqrt{p^2 + m_c^2} }$, the velocity $v/c$ of 
the quark in the $c \bar c$ cms-frame and R(0) radial wave function 
at the origin.

\subsection{Results}

In Table 1 below we show an example of our results for $F(0,0)$.
In Ref.\cite{BGPSS2019} we showed also results for $\eta_c(2S)$.

\begin{table}[h]
\centering
\caption{Transition form factor $|F(0,0)|$ for 
$\eta_{c}(1S)$ at $Q_1^2=Q_2^2 =$0. }
\begin{tabular}{l c c c c} 
\hline
\hline
  potential type &  $m_c$ \, [GeV]  &  $|F(0,0)| \, [\rm{GeV}^{-1}]$
  &  $\Gamma_{\gamma \gamma}$\,[keV]& $f_{\eta_c} $[GeV]\\
\hline
 harmonic oscillator & 1.4   & 0.051 & 2.89 & 0.2757\\
 logarithmic         & 1.5   & 0.052 & 2.95 & 0.3373\\
 power-like          & 1.334  & 0.059 & 3.87 & 0.3074\\  
 Cornell             & 1.84  & 0.039 & 1.69 & 0.3726\\
 Buchm\"uller-Tye    & 1.48  & 0.052 & 2.95 & 0.3276\\
 \hline
 experiment & - & 0.067 $\pm$ 0.003 [1] & 5.1
 $\pm$ 0.4 [1]& 0.335  $\pm$ 0.075 [2]\\
\hline
\hline
\end{tabular}
\end{table}

\begin{figure}
\centering
\includegraphics[width=0.48\linewidth]{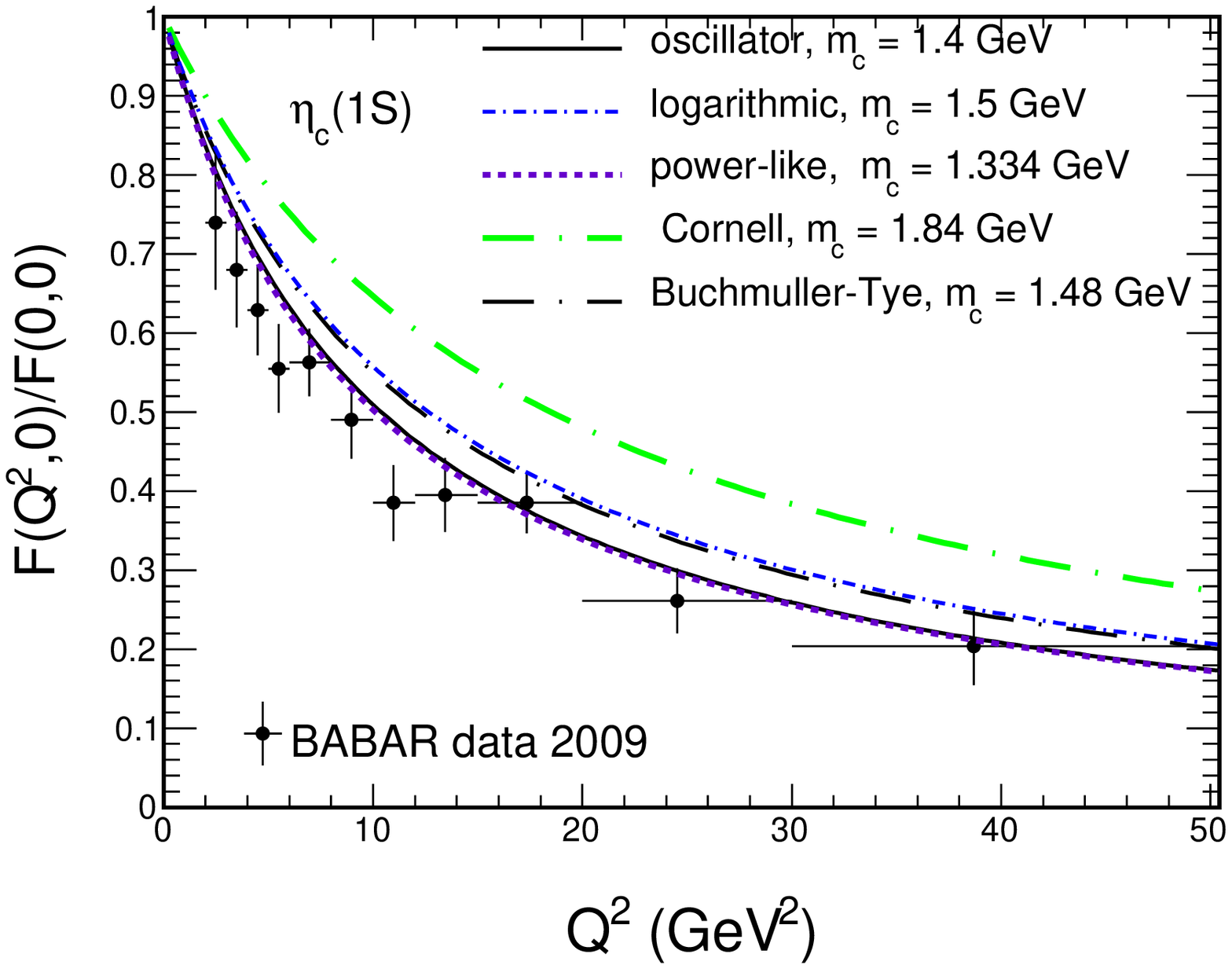}
\includegraphics[width=0.48\linewidth]{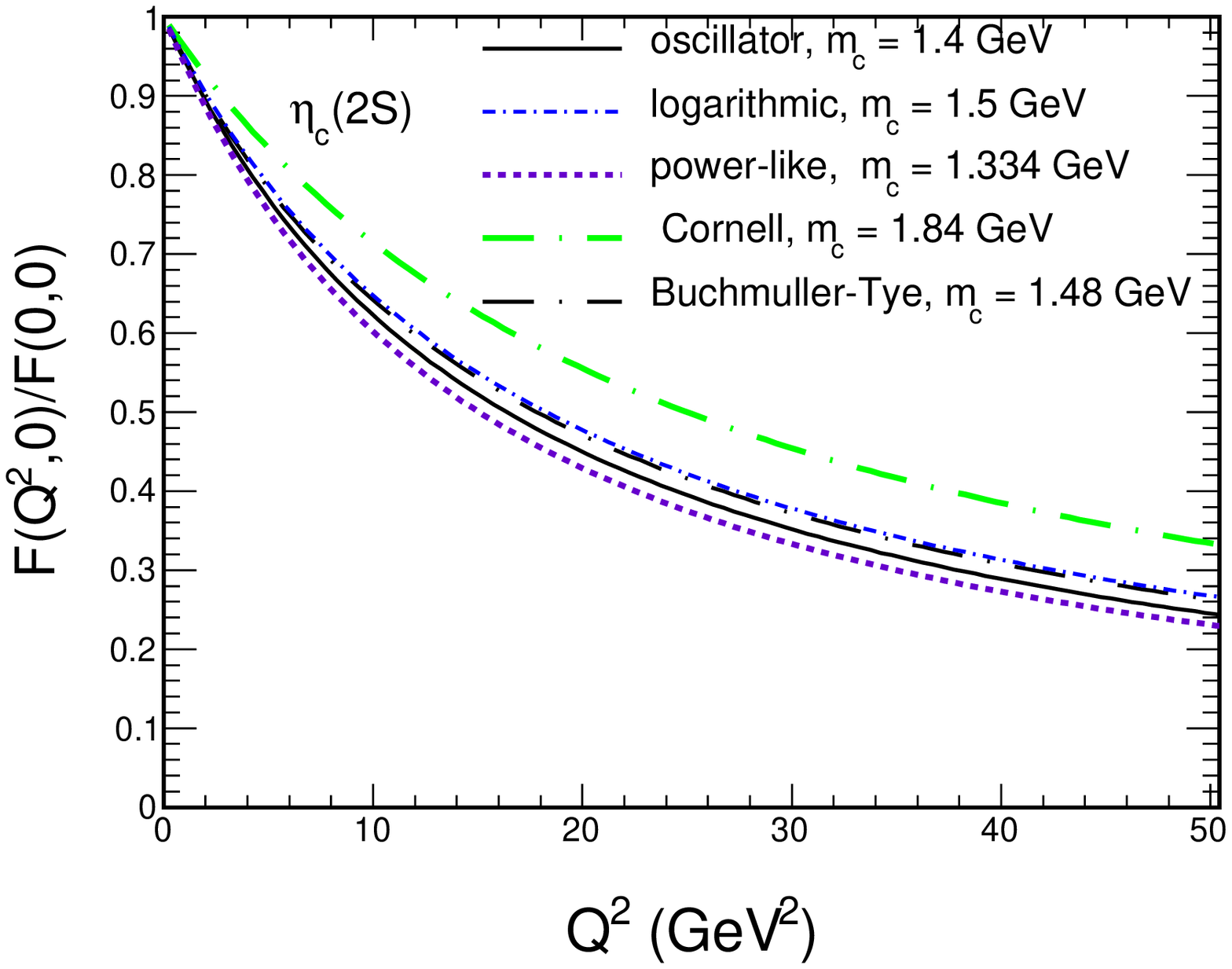}
\caption{ Normalized transition form factor 
$\tilde{F}(Q^2,0)$ as a function of photon virtuality~$Q^2$. The BaBar 
data are shown for comparison (see J.~P.~Lees {\it et al.} 
\cite{Lees:2010de} [BaBar Collaboration]).
}
\label{fig:F_Q2}
\end{figure}

Let us start presentation of our results for transition form factor
for one real and one virtual photon. Such objects are measured
for single-tagged $e^+ e^- \to e^+ e^- \eta_c(1S)$ reaction, i.e.
when only one scattered electron/positron is measured.
In Fig.\ref{fig:F_Q2} we show results of our calculations for different
wave functions (potentials) for $\eta_c(1S)$. For comparison we show
also experimental form factor extracted by the Babar collaboration
\cite{Lees:2010de}. The theoretical results depend on the potential used.
For some models the agreement is better than for the other models.
As discussed in \cite{BGPSS2019}, the results depend rather on the mass
of the charm quark/antiquark and much less on particular form of the
$c \bar c$ potential. 

\begin{figure}
\centering\includegraphics[width=0.39\linewidth]{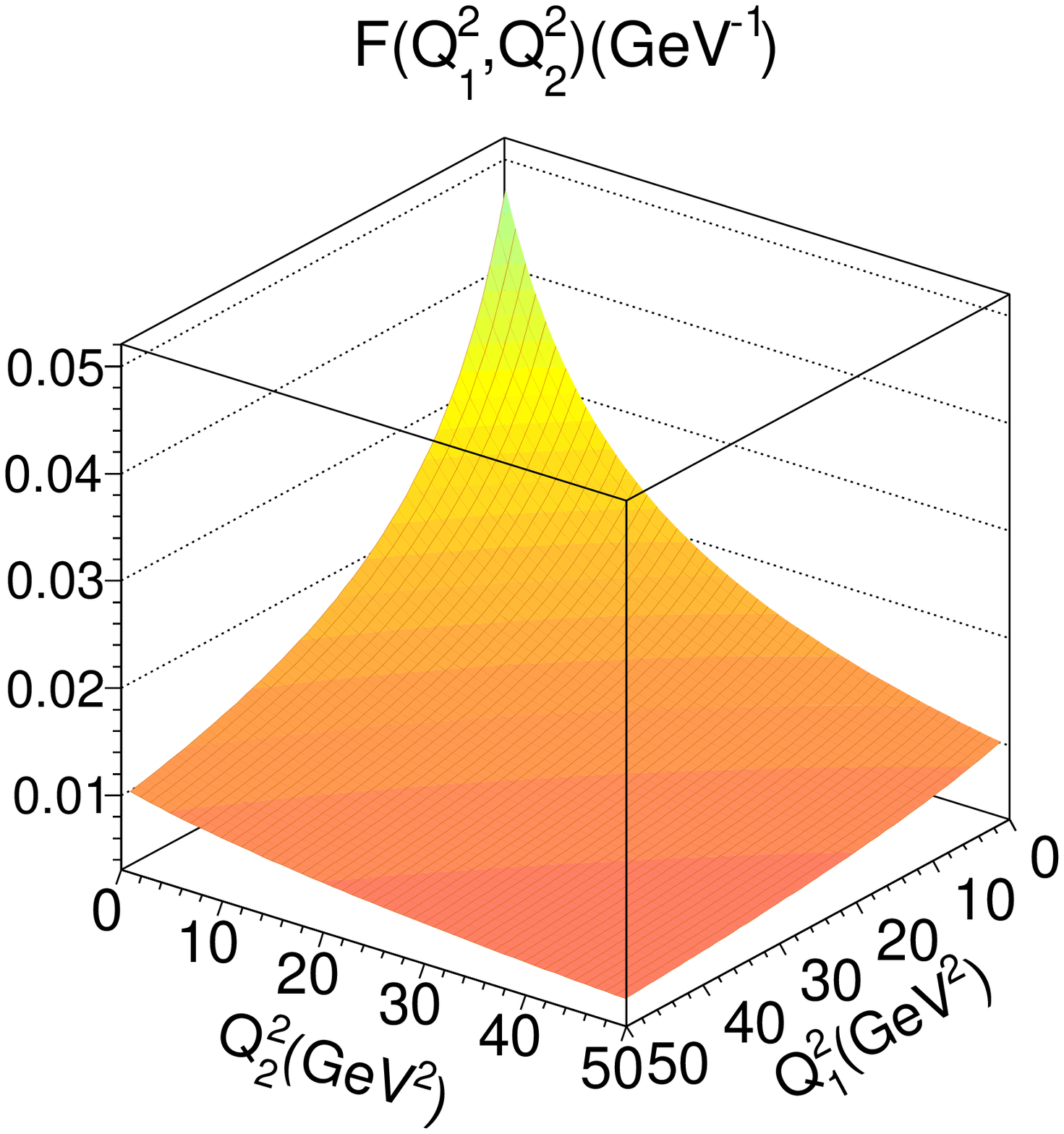}
\centering\includegraphics[width=0.39\linewidth]{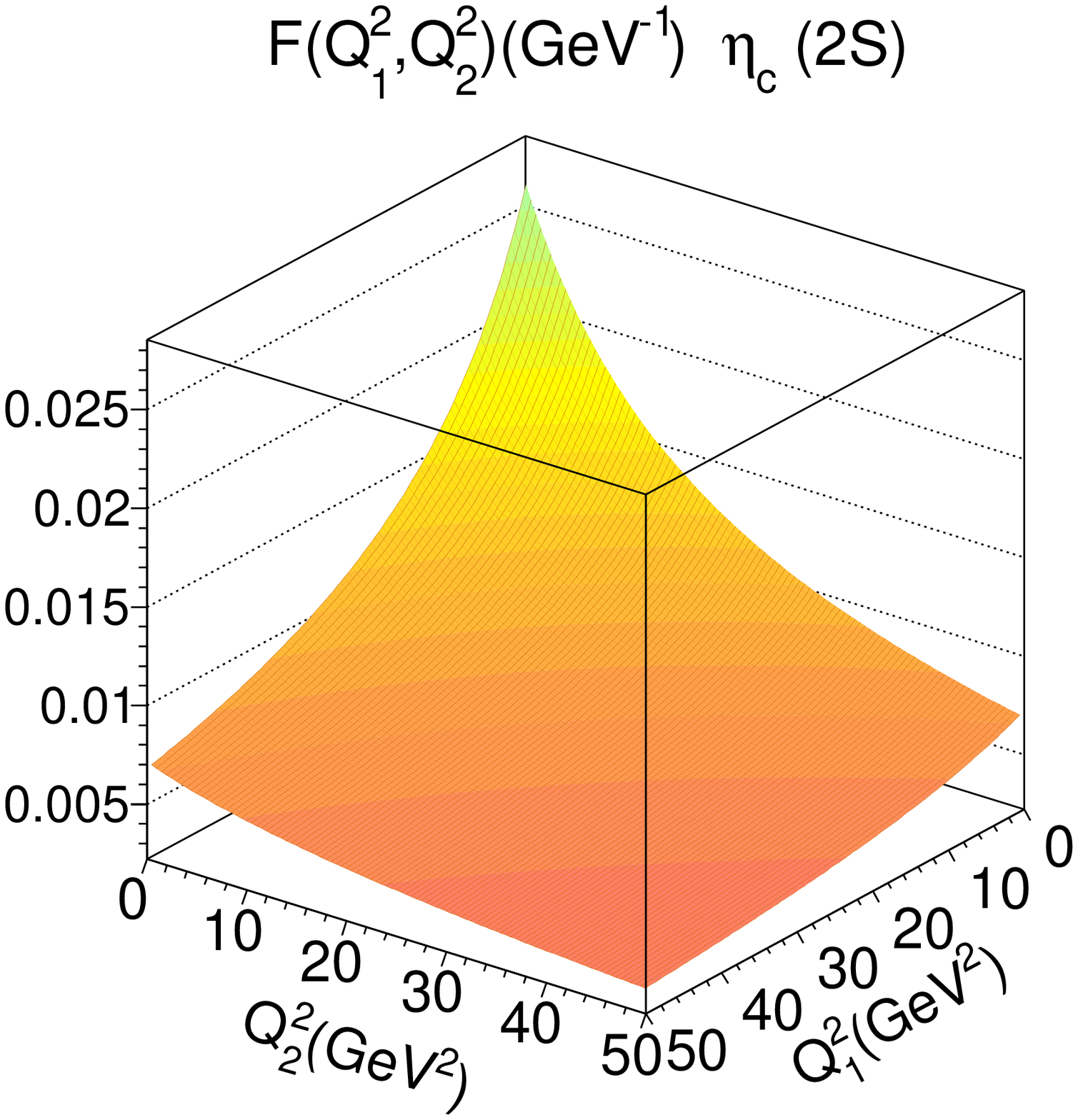}
\caption{ Transition form factor for $\eta_c(1S)$ and $\eta_c(2S)$ 
for Buchm\"uller-Tye potential.
The~$F(Q_{1}^{2}, Q_{2}^{2})$ should obey Bose symmetry.}
\label{fig:Q12Q22_dependence}
\end{figure}

In Fig.\ref{fig:Q12Q22_dependence} we show the dependence of the
transition form factors on both photon virtualities
for $\eta_c(1S)$ (left panel) and $\eta_c(2S)$ (right panel)
as an example for the Buchm\"uller-Tye potential. Such distributions 
were shown in \cite{BGPSS2019} for the first time.

\begin{figure}
\centering
\includegraphics[width=0.39\linewidth]{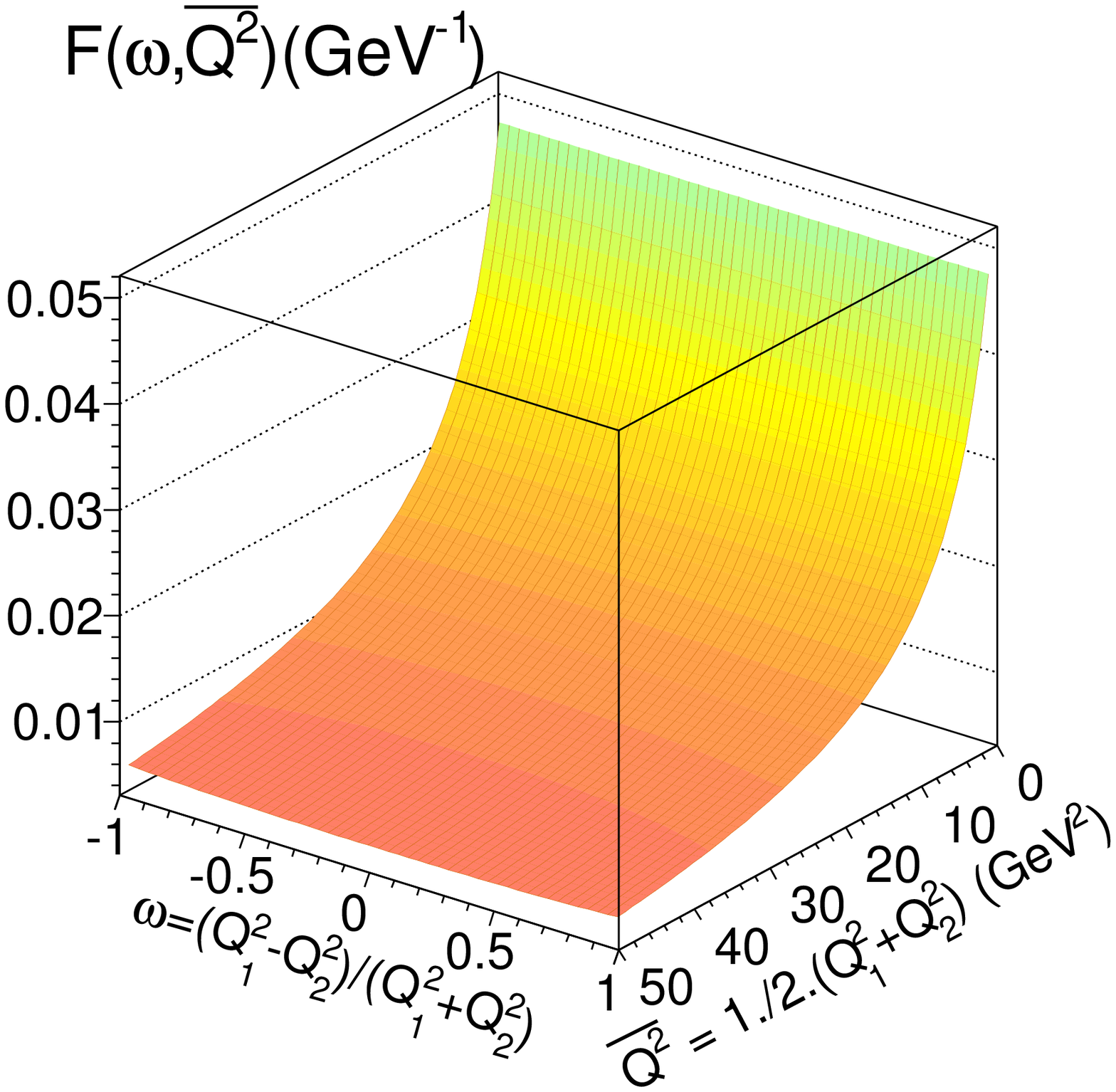}
\includegraphics[width=0.39\linewidth]{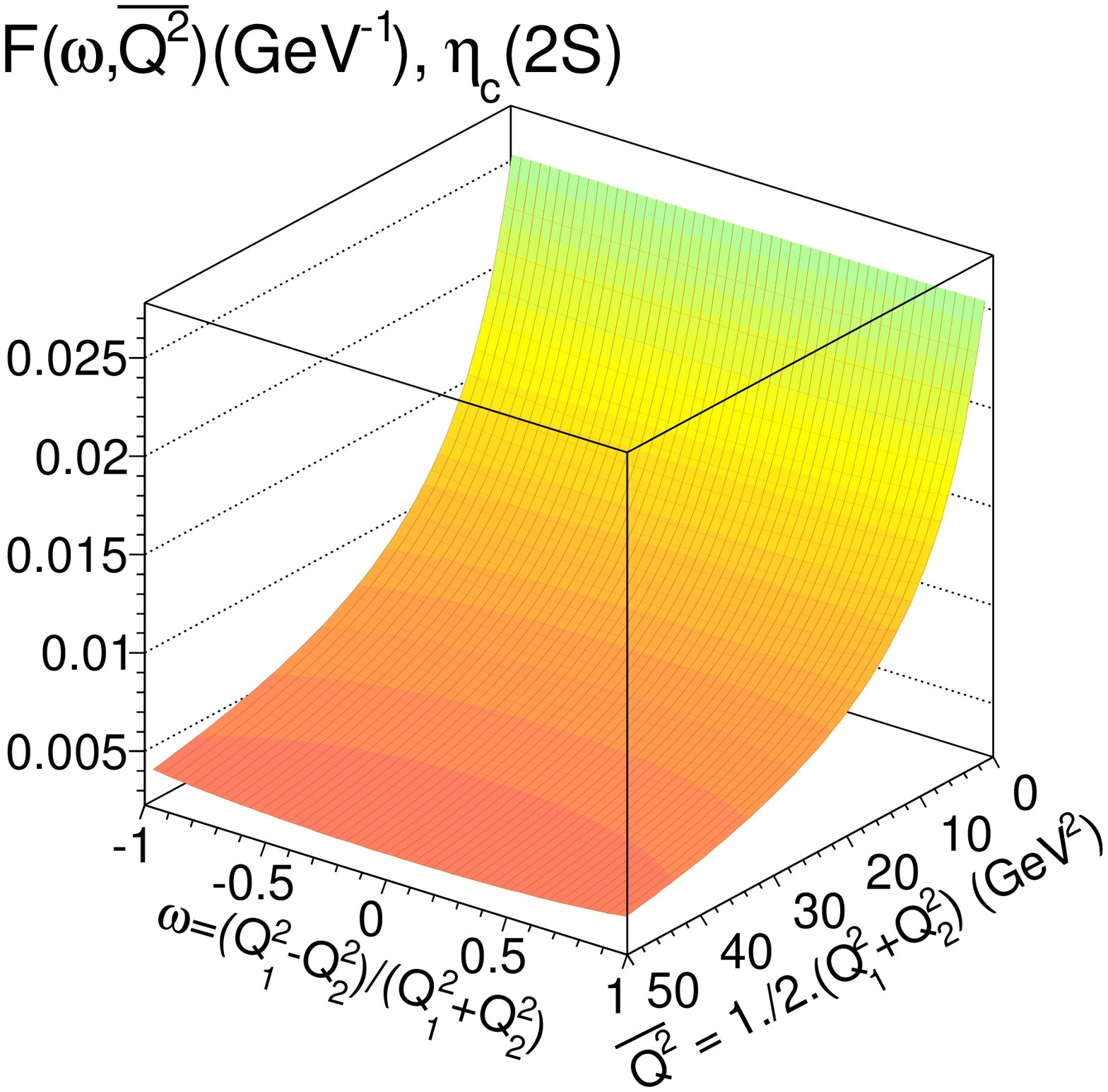}
\caption{
The $\gamma^*\gamma^*\rightarrow \eta_{c}$ (1S) and
             $\gamma^*\gamma^*\rightarrow \eta_{c}$ (2S)
form factor as a function of ($\omega,\bar{Q^2}$)
for the Buchm\"uller-Tye potential for illustration.
}
\label{fig:omegaQ2a_dependence}
\end{figure}

In Fig.\ref{fig:omegaQ2a_dependence} we show the form factors
in slightly different representation:
\begin{eqnarray}
\omega=\frac{Q_1^2-Q_2^2}{Q_1^2+Q_2^2}\, \,\,\,\mbox{and}\,\,\,\bar{Q}^2 = \frac{Q_1^2+Q_2^2}{2}\, . \nonumber
\end{eqnarray}
We observe scaling in the $\omega$ variable. 

\begin{figure}[h]
\centering\includegraphics[width=0.48\linewidth]{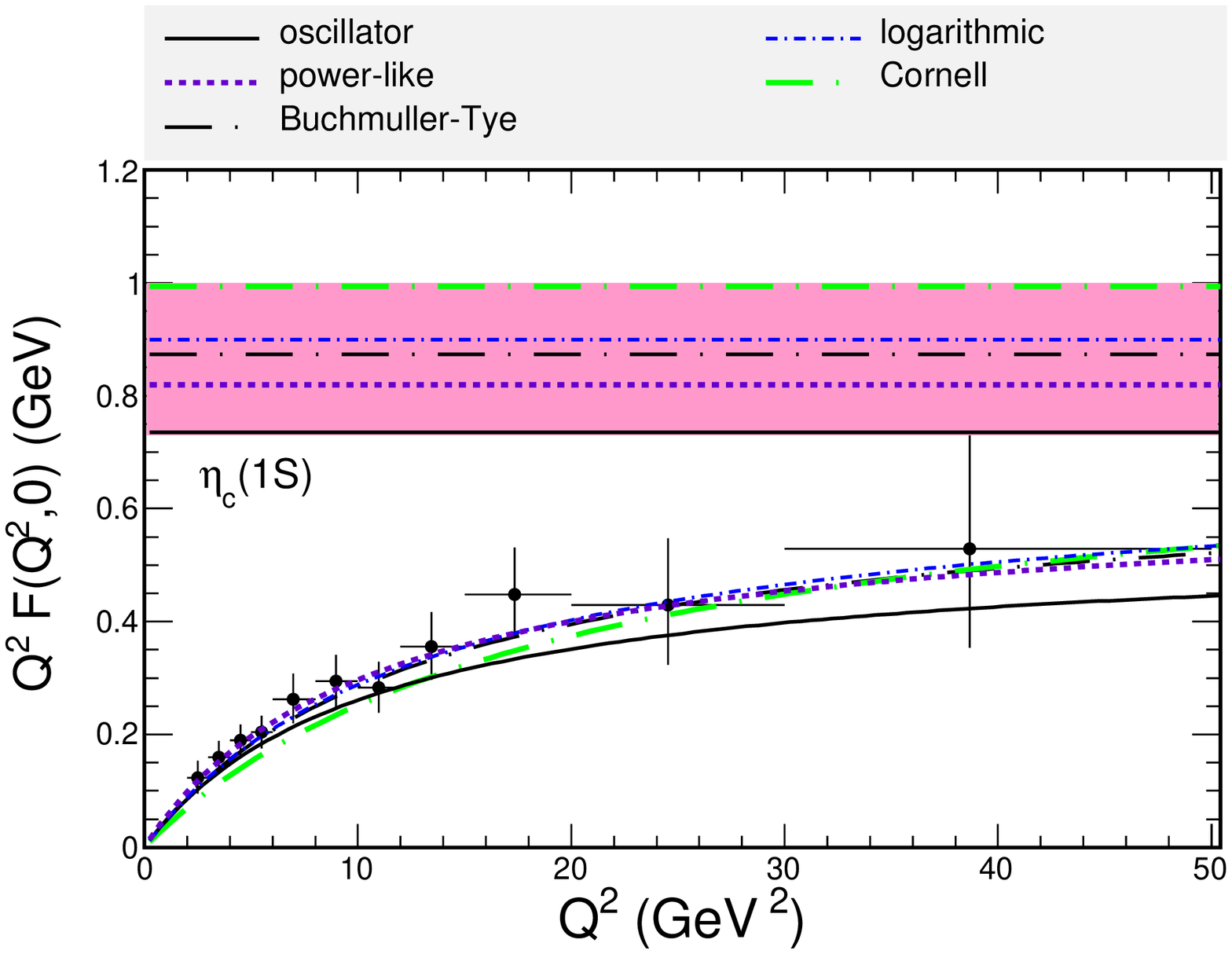}
\centering\includegraphics[width=0.48\linewidth]{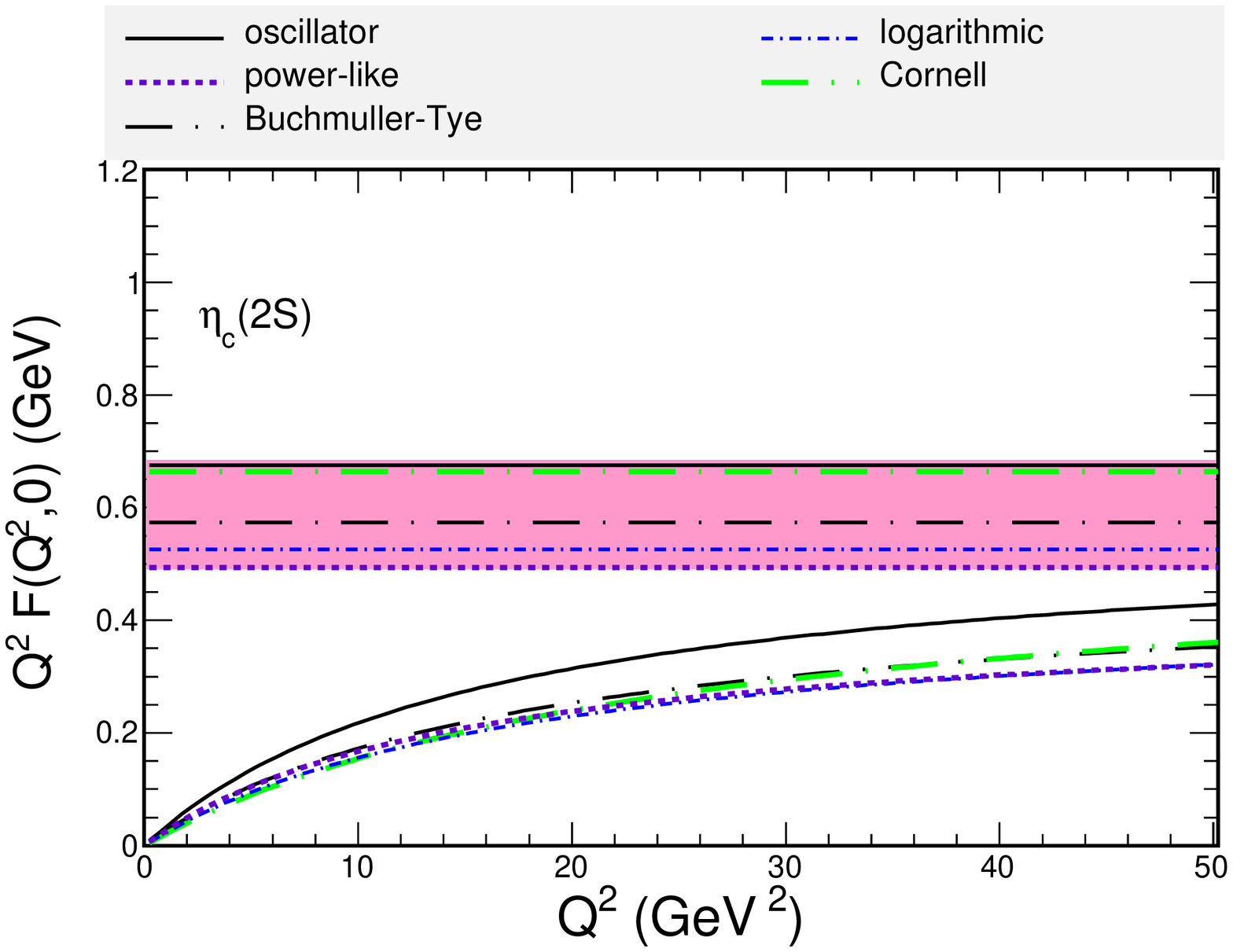}
\caption{$Q^2F(Q^2,0)$ as a function of photon virtuality~$Q^2$.
The horizontal lines $\frac{8}{3}f_{\eta_{c}}$ are shown for reference.}
\label{fig:Q2F}
\end{figure}

The convergence of $Q^2 F(Q^2)$ to its asymptitic value is shown in
Fig.\ref{fig:Q2F} for different potentials used in \cite{BGPSS2019}.
Even at $Q^2 \sim$ 30 GeV$^2$ our results is very far for the asymptotic
value (different for different wave functions). The effect of RGE was
discussed in \cite{BGPSS2019} and was shown to be very slow.

\section{Inclusive production of $\eta_c$ quarkonia in
    proton-proton collisions}

\subsection{Theoretical approach}

\begin{figure}
    \centering
    \includegraphics[width=0.6\textwidth]{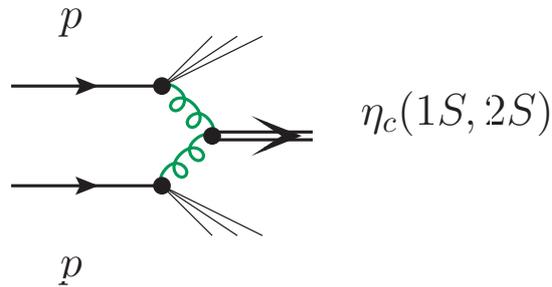}
    \caption{Generic diagram for the inclusive process of 
$\eta_{c}$(1S) or $\eta_{c}$(2S) production in $pp$ scattering 
via two gluons fusion. }
    \label{fig:diagram_etac}
\end{figure}

The diagram shown in Fig.\ref{fig:diagram_etac} illustrates the situation
adequate for the $k_T$-factorization calculations used in 
Ref.\cite{BPSS2020}.

The inclusive cross section for $\eta_c$-production via the $2 \to 1$ 
gluon-gluon fusion mode is obtained from
\begin{eqnarray}
d\sigma = \int {dx_1 \over x_1} \int {d^2 \bq_1 \over \pi \bq_1^2} 
{\cal{F}}(x_1,\bq_1^2)
\int {dx_2 \over x_2} \int {d^2 \bq_2 \over \pi \bq_2^2}  
{\cal{F}}(x_2,\bq_2^2) 
{1 \over 2 x_1 x_2 s} \overline{|{\cal{M}}|}^2 \, d\Phi(2 \to 1).
\end{eqnarray}
The unintegrated gluon distributions are normalized such, that in 
the DGLAP-limit
\begin{eqnarray}
{\cal{F}}(x,\bq^2) = {\partial x g(x,\bq^2) \over \partial \log \bq^2}
\; .
\end{eqnarray}
Let us denote the four-momentum of the $\eta_c$ by $P$. It can
be parametrized as:
\begin{eqnarray}
P = (P_+,P_-,\bP) = ({m_\perp \over \sqrt{2}} e^y, {m_\perp \over \sqrt{2}} e^{-y}, \bP) \, , 
\end{eqnarray}

We therefore obtain for the inclusive cross section
\begin{eqnarray}
{d \sigma \over dy d^2\bP} = \int {d^2 \bq_1 \over \pi \bq_1^2} 
{\cal{F}}(x_1,\bq_1^2) \int {d^2 \bq_2 \over \pi \bq_2^2}  {\cal{F}}(x_2,\bq_2^2) \, \delta^{(2)} (\bq_1 + \bq_2 - \bP ) \, {\pi \over (x_1 x_2 s)^2} \overline{|{\cal{M}}|}^2 ,
\end{eqnarray}
where the momentum fractions $x_{1,2}$ of gluons are
\begin{eqnarray}
x_1 = {m_\perp \over \sqrt{s}} e^y \, , \, x_2 = {m_\perp \over \sqrt{s}} e^{-y} .
\end{eqnarray}
The off-shell color singlet matrix element 
is written in terms of the Feynman amplitude as:
\begin{eqnarray}
{\cal{M}}^{ab} = {q_{1 \perp}^\mu q_{2\perp}^\nu \over |\bq_1| |\bq_2|}{\cal{M}}^{ab}_{\mu \nu}  = {q_{1+} q_{2-} \over |\bq_1| |\bq_2|} n^+_\mu n^-_\nu {\cal{M}}^{ab}_{\mu \nu} = {x_1 x_2 s \over 2 |\bq_1| |\bq_2| } n^+_\mu n^-_\nu {\cal{M}}^{ab}_{\mu \nu}  \, .
\end{eqnarray}
Then, we obtain for the cross section
\begin{eqnarray}
{d \sigma \over dy d^2\bP} = \int {d^2 \bq_1 \over \pi \bq_1^4} 
{\cal{F}}(x_1,\bq_1^2) 
\int {d^2 \bq_2 \over \pi \bq_2^4}  
{\cal{F}}(x_2,\bq_2^2) \, 
\delta^{(2)} (\bq_1 + \bq_2 - \bP ) \, 
{\pi \over 4} \overline{|n^+_\mu n^-_\mu{\cal{M}}_{\mu \nu}|}^2 ,
\end{eqnarray}
It is related to the $\gamma^* \gamma^* \eta_c$ transition form factor 
through the relation
\begin{eqnarray}
F(Q_1^2,Q_2^2) = e_c^2 \sqrt{N_c} \, I(\bq_1^2,\bq_2^2) \, .
\end{eqnarray}
The vector product $[\bq_1,\bq_2]$ is defined as
\begin{eqnarray}
[\bq_1,\bq_2] = q_1^x q_2^y - q_1^y q_2^x = |\bq_1| |\bq_2| 
\sin(\phi_1 - \phi_2) \, .
\end{eqnarray}

Then, the averaged matrix element squared becomes
\begin{eqnarray}
\overline{|n^+_\mu n^-_\mu{\cal{M}}_{\mu \nu}|}^2
&=& 16 \pi^2 \alpha_S^2 {1 \over 4} {1 \over N_c} |[\bq_1,\bq_2] \, I(\bq_1^2,\bq_2^2)|^2   {1 \over (N_c^2 -1)^2} \sum_{a,b} \delta^{ab} \delta^{ab} \nonumber \\
&=& 4 \pi^2 \alpha_S^2 {1 \over N_c (N_c^2 -1)}  |[\bq_1,\bq_2] \, I(\bq_1^2,\bq_2^2)|^2 
\end{eqnarray}

This leads to our final result:
\begin{eqnarray}
{d \sigma \over dy d^2\bP} = \int {d^2 \bq_1 \over \pi \bq_1^4} 
{\cal{F}}(x_1,\bq_1^2) 
\int {d^2 \bq_2 \over \pi \bq_2^4}  
{\cal{F}}(x_2,\bq_2^2) 
\, \delta^{(2)} (\bq_1 + \bq_2 - \bP ) \, 
{\pi^3 \alpha_S^2 \over N_c
  (N_c^2-1)} |[\bq_1,\bq_2] \, I(\bq_1^2,\bq_2^2)|^2  .
\nonumber 
\end{eqnarray}

In real calculation we take $\mu_F^2 = m_T^2$ and for 
renormalization scale(s)
\begin{equation}
\alpha_s^2 \to 
\alpha_s(max(m_t^2,q_{t,1}^2))
\alpha_s(max(m_t^2,q_{t,2}^2))  \; .
\label{alpha_s}
\end{equation}
From the proportionality of the $g^* g^* \eta_c$ and 
$\gamma^* \gamma^* \eta_c$ vertices to the leading order (LO), 
we obtain, that at LO:
\begin{eqnarray}
\label{eq:LO_gluon}
\Gamma_{\rm{LO}}(\eta_c \to gg) = {N_c^2 -1 \over 4 N_c^2} \, {1 \over e_c^4} \,
\Big( {\alpha_s \over \alpha_{\rm em}} \Big)^2 \, 
\Gamma_{\rm{LO}}(\eta_c \to \gamma \gamma) \, ,
\end{eqnarray}
where the LO $\gamma \gamma$ width is related to the
transition form factor for vanishing virtualities through
\begin{equation}
\label{eq:L0_photon}
    \Gamma_{\rm{LO}}(\eta_c \to \gamma \gamma) = 
   {\pi \over 4} \alpha^2_{\rm em}
    M^3_{\eta_c} |F(0,0)|^2 \, .
\end{equation}
At NLO, the expressions for the widths read 
(see \cite{Lansberg:2006dw})
\begin{eqnarray}
\label{eq:NLO}
\Gamma(\eta_c \to \gamma \gamma) &=& \Gamma_{\rm{LO}}(\eta_c \to \gamma
\gamma) \, \Big( 1 - {20 - \pi^2 \over 3} 
{\alpha_s \over \pi} \Big)\,, \nonumber \\
\Gamma(\eta_c \to g g) &=& \Gamma_{\rm{LO}}(\eta_c \to g g) \, 
\Big( 1 + 4.8 \, {\alpha_s \over \pi} \Big). 
\end{eqnarray}
%
\begin{table}[]
    \centering
    \caption{Total decay widths as well as $|F(0,0)|$ obtained from
      $\Gamma_{tot}$ using the next-to-leading order approximation.}
    \begin{tabular}{c|c|c}
    
    \hline
    \hline
     &Experimental values & Derived from Eq.(\ref{eq:NLO}) \\
      
      & $\Gamma_{tot}$ (MeV)  &$|F(0,0)|_{gg} [GeV^{-1}]$\\ 
                      \hline
    $\eta_{c}(1S)$     & 31.9$\pm$0.7 &0.119$\pm$0.001\\
   $\eta_{c}(2S)$     & 11.3$\pm$3.2$\pm$2.9 &0.053$\pm$0.010 \\
    \hline
    \hline
    \end{tabular}
    \label{table:width_tot}
\end{table}

We use a few different UGDs which are available from the literature,
e.g. from the TMDLib package (see \cite{Hautmann:2014kza}) 
or the CASCADE Monte Carlo code (see \cite{Jung:2010si}).

\begin{enumerate}

    \item Firstly we use a glue constructed according to the
      prescription initiated in Kimber et al.\cite{Kimber:2001sc}
      and later  updated in Martin et al.\cite{Martin:2009ii}), 
     which we label below as ``KMR''. 
    It uses as an input the collinear gluon distribution from 
    Harland-Lang et al.\cite{Harland-Lang:2014zoa}. 

    \item Secondly, we employ two UGDs obtained by
      Kutak \cite{Kutak:2014wga}. There are two
      versions of this UGD. Both introduce a hard scale dependence via a
      Sudakov form factor into solutions of a small-$x$ evolution equation.
    The first version uses the solution of a linear, BFKL 
    evolution with
    a resummation of subleading terms and is denoted by 
    ''Kutak (linear)''. The second UGD, denoted as 
    ``Kutak (nonlinear)'' uses
    instead a nonlinear evolution equation of Balitsky-Kovchegov
    type. Both of the Kutak's UGDs \cite{Kutak:2014wga} can be applied 
    only in the small-$x$ regime, $x < 0.01$.

    \item The third type of UGD has been obtained by
      Hautmann and Jung from a description of 
      precise HERA data on deep inelastic structure function by a 
      solution of the CCFM evolution equations. We use ``Set 2''.
     from \cite{Hautmann:2013tba}.

\end{enumerate}

\subsection{Results}

\begin{figure}
    \centering
    \includegraphics[width=0.45\textwidth]{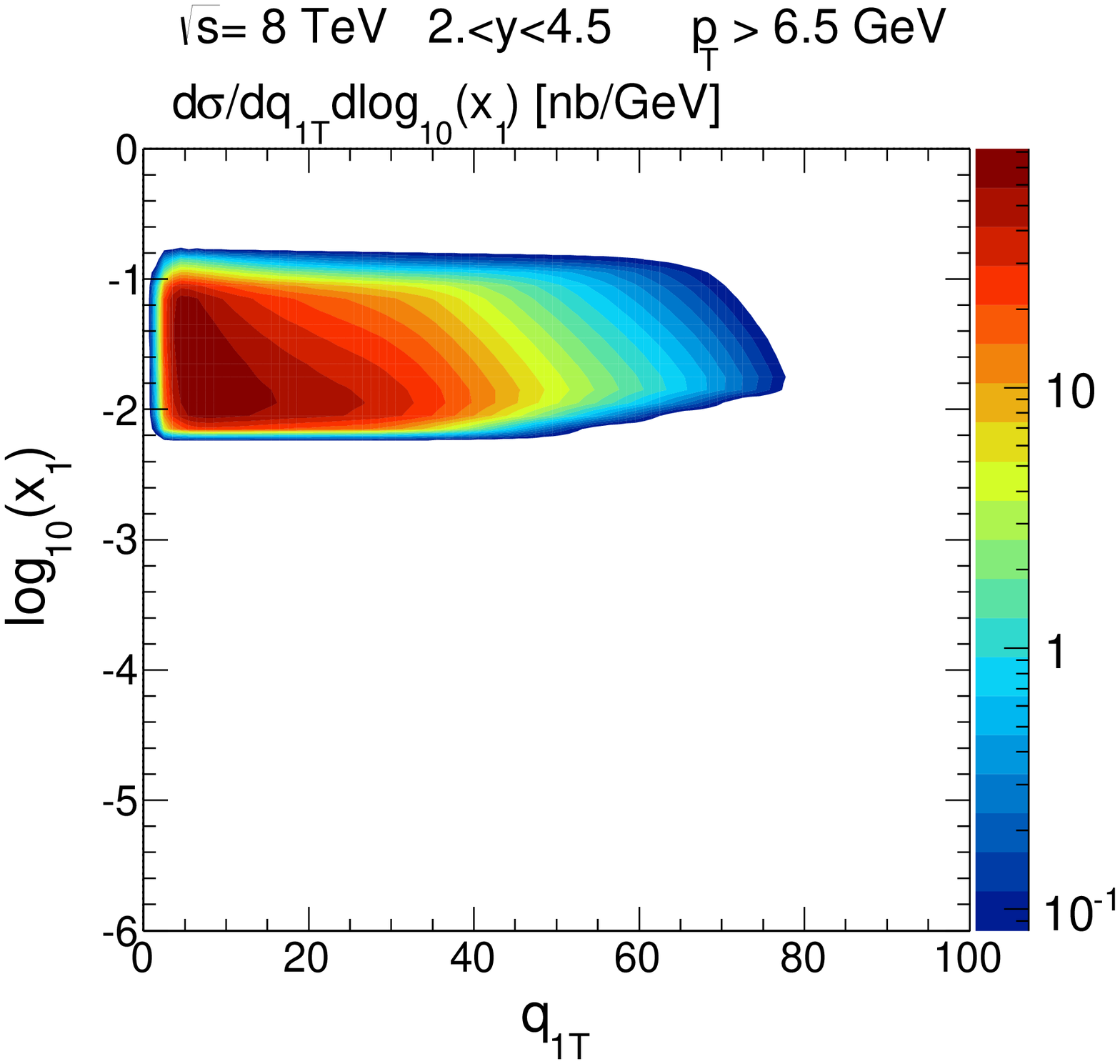}
    \includegraphics[width=0.45\textwidth]{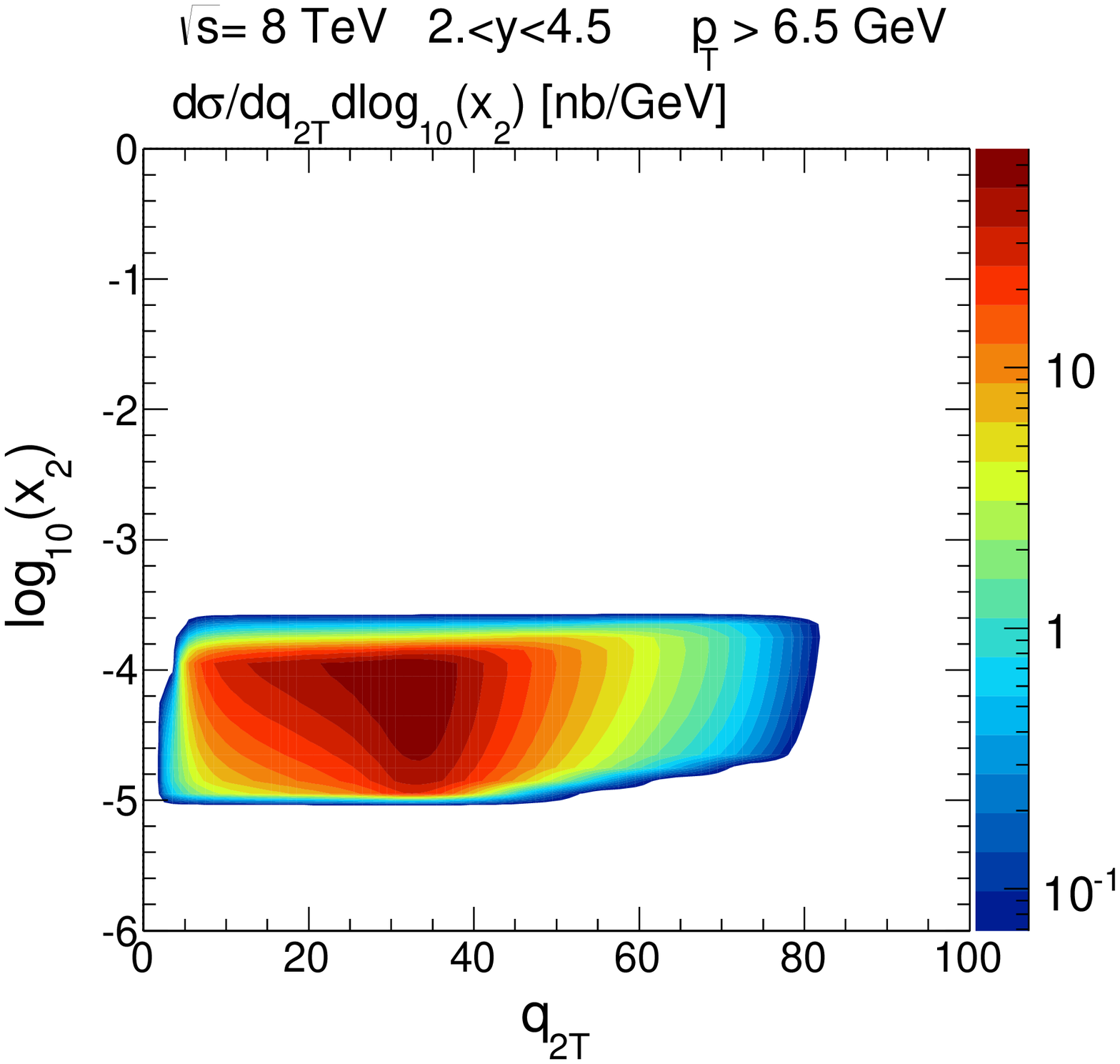}
    \caption{Two-dimensional distributions in $(x_1, q_{1T})$
(left panel) and in $(x_2, q_{2T})$ (right panel) for $\eta_c(1S)$
production for $\sqrt{s}$ = 8 TeV.
In this calculation the KMR UGD was used for illustration.
} 
    \label{fig:dsig_dxdkt}
\end{figure}

In Fig.\ref{fig:dsig_dxdkt} we show the cross section distributions
for $(x_1, q_{1T})$ (left panel) and $(x_2, q_{2T})$ (right panel).
For the LHCb kinematics the two distributions are not identical:
$x_1 \gg x_2$ and on average $q_{1T} < q_{2T}$.

\begin{figure}
    \centering
    \includegraphics[width=0.45\textwidth]{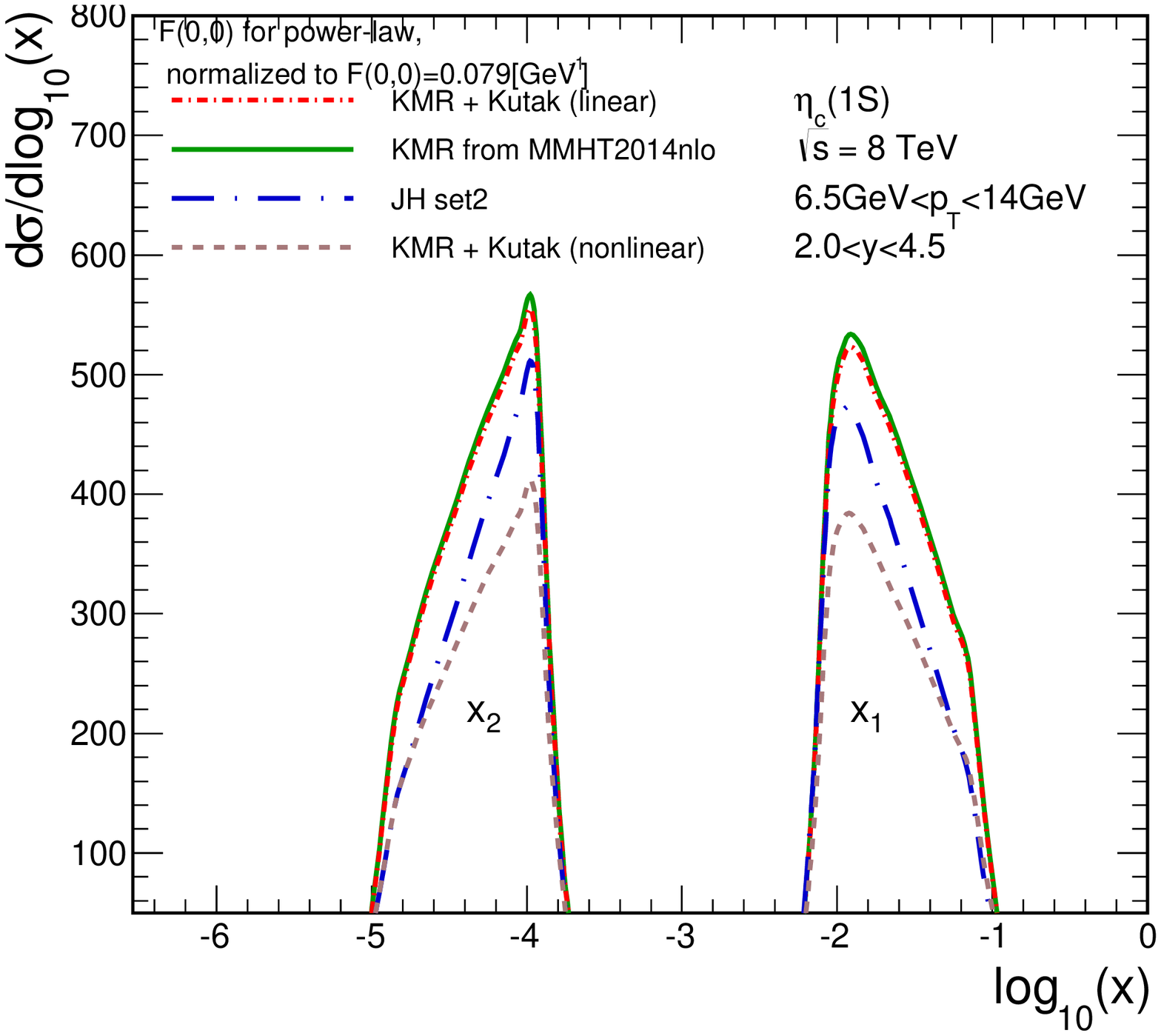}
    \includegraphics[width=0.45\textwidth]{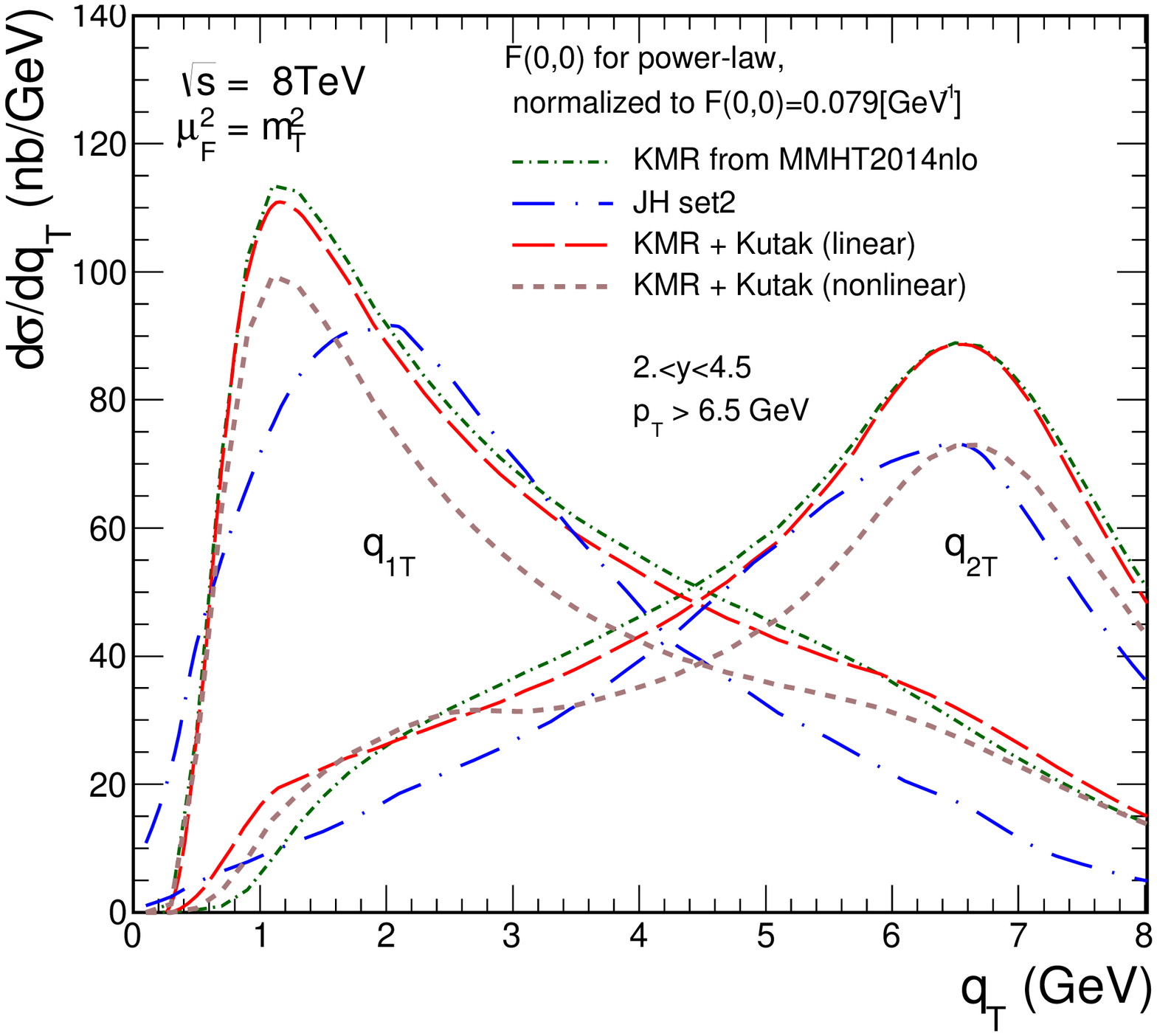}
    \caption{Distributions in  $\log_{10}(x_1)$ or $\log_{10}(x_2)$ 
(left panel) and distributions in $q_{1T}$ or $q_{2T}$    
(right panel) for the LHCb kinematics. Here the selected UGDs were used
in our calculations. Here we show an example for $\sqrt{s}$ = 8 TeV. }
    \label{fig:projections}
\end{figure}

The projections on longitudinal mementum fraction and gluon transverse
momentum squared are shown in the left and right panels of
Fig.\ref{fig:projections}.

\begin{figure}[!h]
    \centering
    \includegraphics[width=0.32\textwidth]{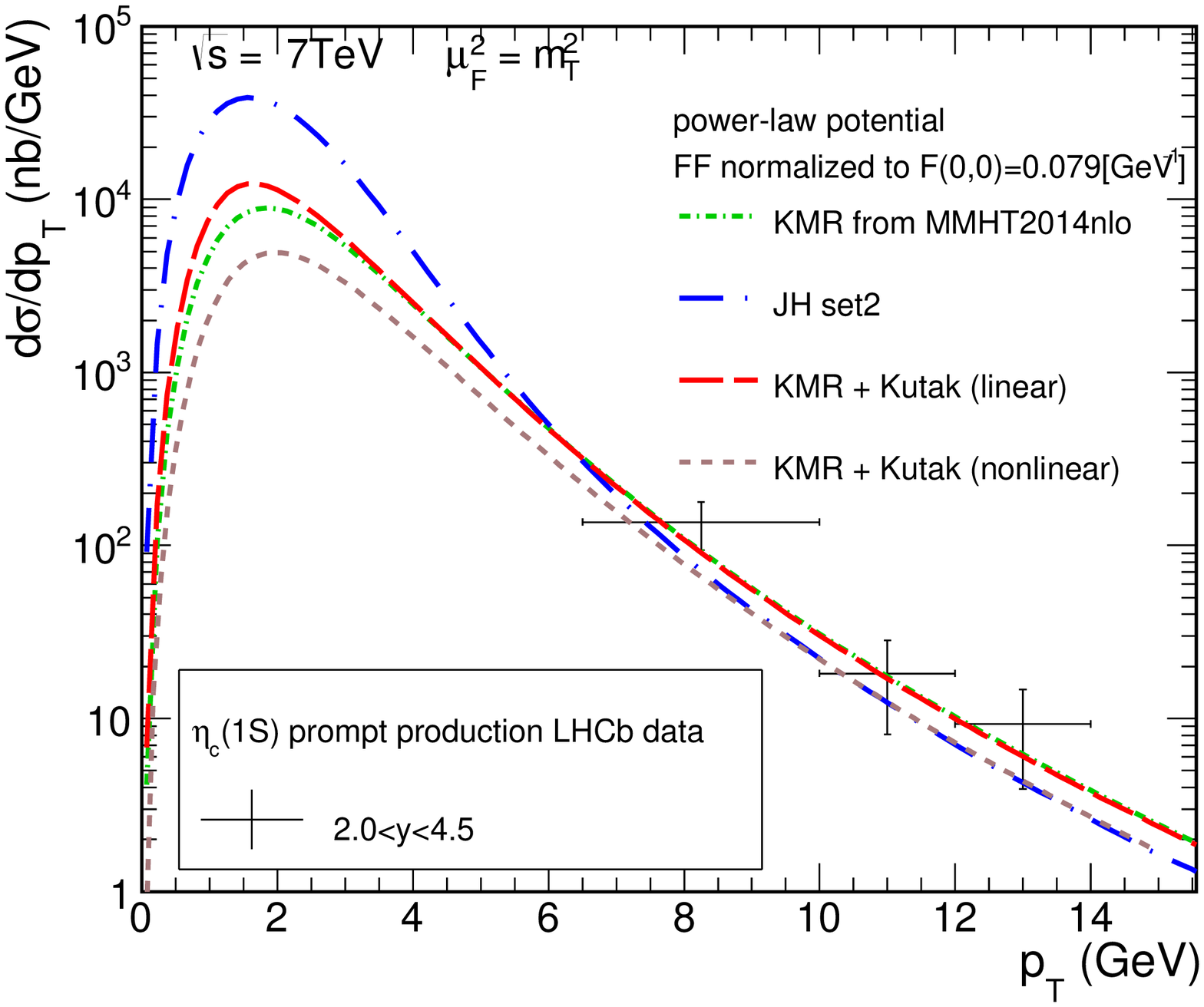}
    \includegraphics[width=0.32\textwidth]{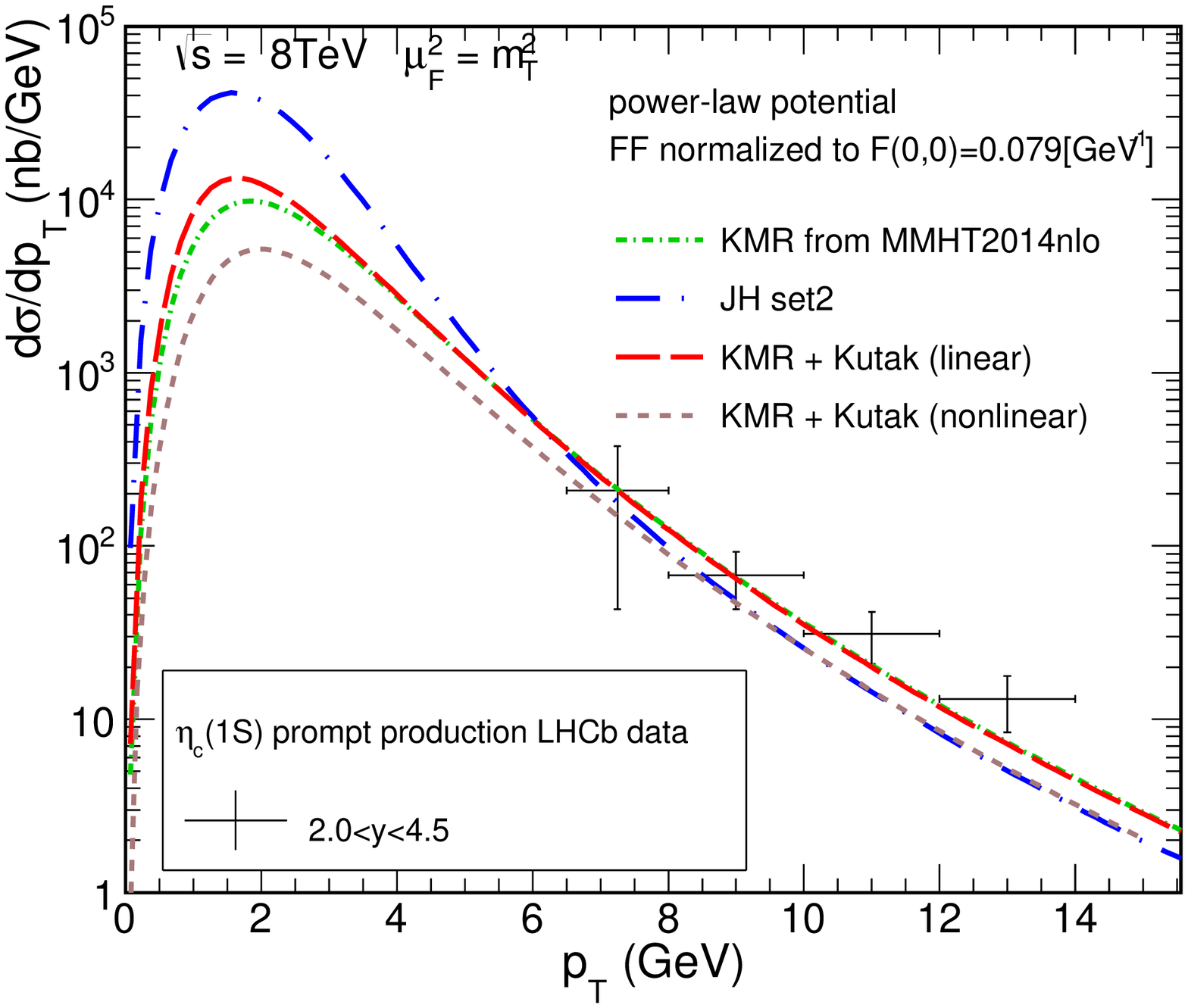}
    \includegraphics[width=0.32\textwidth]{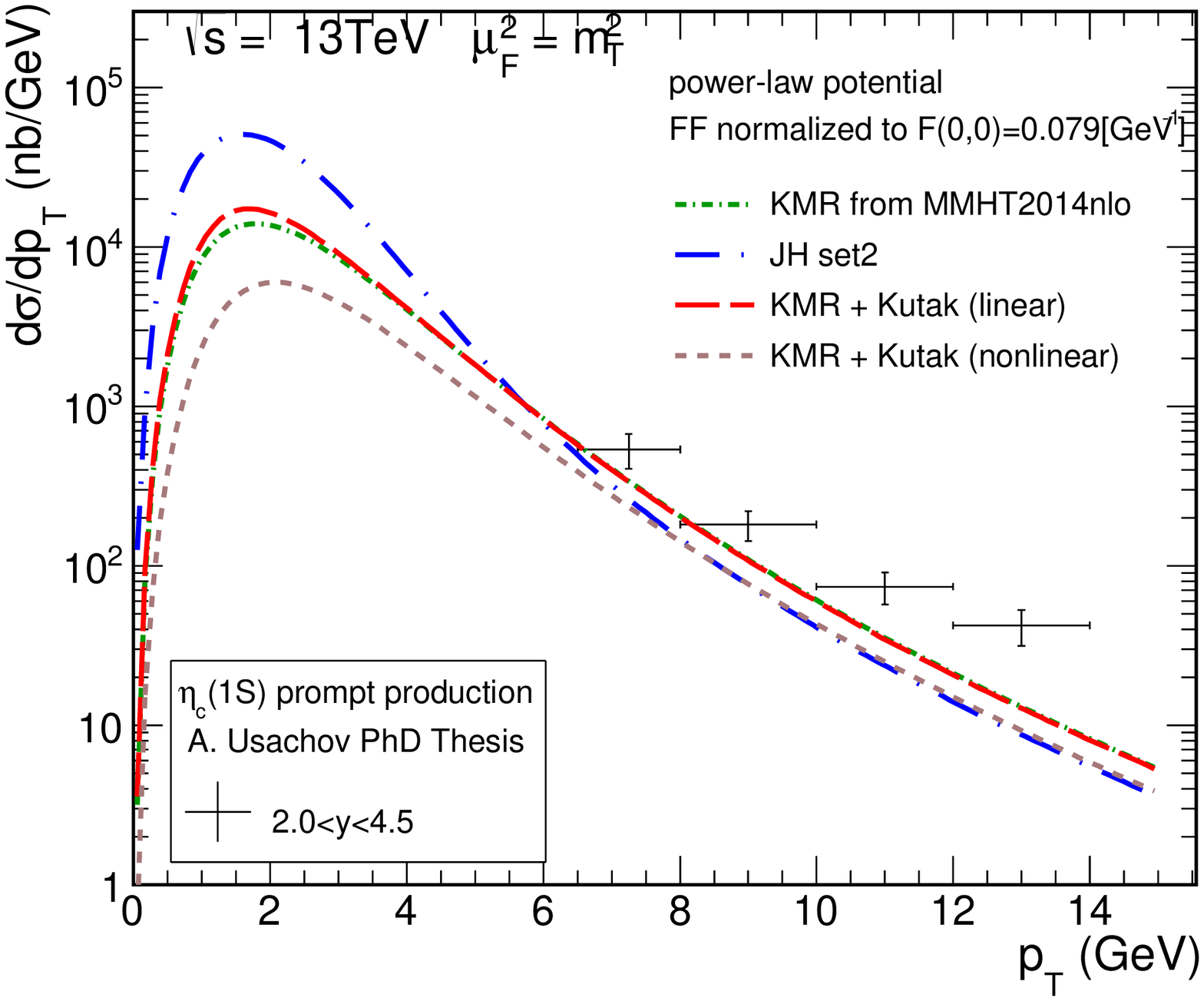}
    \caption{Differential cross section as a function of transverse 
      momentum for prompt $\eta_{c}(1S)$ production compared with 
      the LHCb data 
      (see \cite{Aaij:2014bga}) for $\sqrt{s} = 7, 8 \, \rm{TeV}$
      and preliminary experimental data (Usachov PhD \cite{Usachov:2019czc}) 
     for $\sqrt{s}$ = 13 TeV.
     Different UGDs were used. Here we used the 
     $g^* g^* \to \eta_c(1S)$ form factor calculated from the power-law 
     potential.}
\label{fig:dsig_dpt_etac1S}    
\end{figure}

Transverse momentum distributions for $\eta_c(1S)$ are shown
in Fig.\ref{fig:dsig_dpt_etac1S} for three different collision energies
for different unintegrated gluon distributions specified in the figure.
The LHCb data points are shown for comparison.
Our theoretical results almost agree with the LHCb data for $\sqrt{s} =$
7 and 8 TeV while at $\sqrt{s} =$ 13 TeV the preliminary experimental
data are above our predictions. We have no idea how to explain the
disagreement.

\begin{figure}[!h]
    \centering
    \includegraphics[width=0.32\textwidth]{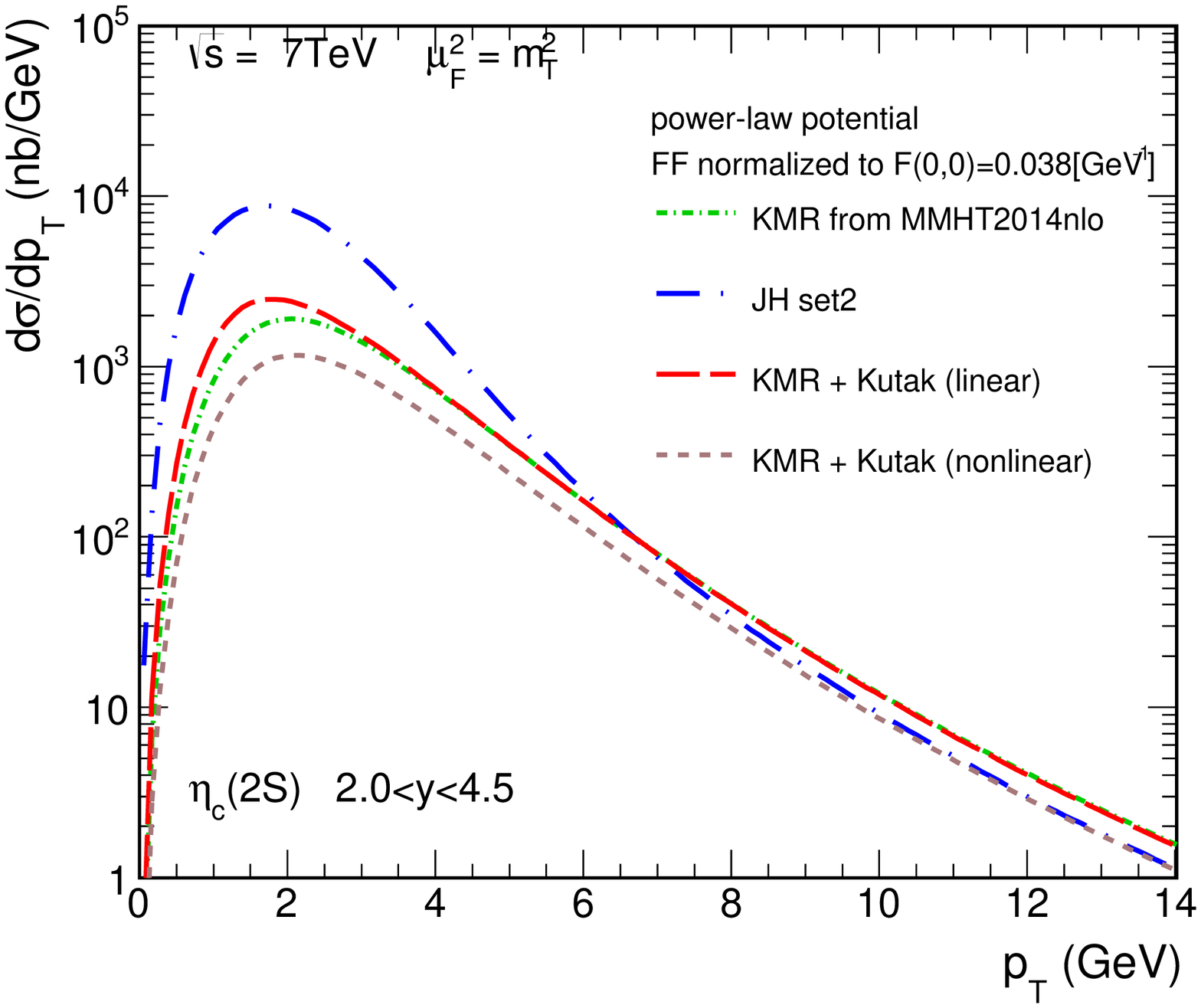}
    \includegraphics[width=0.32\textwidth]{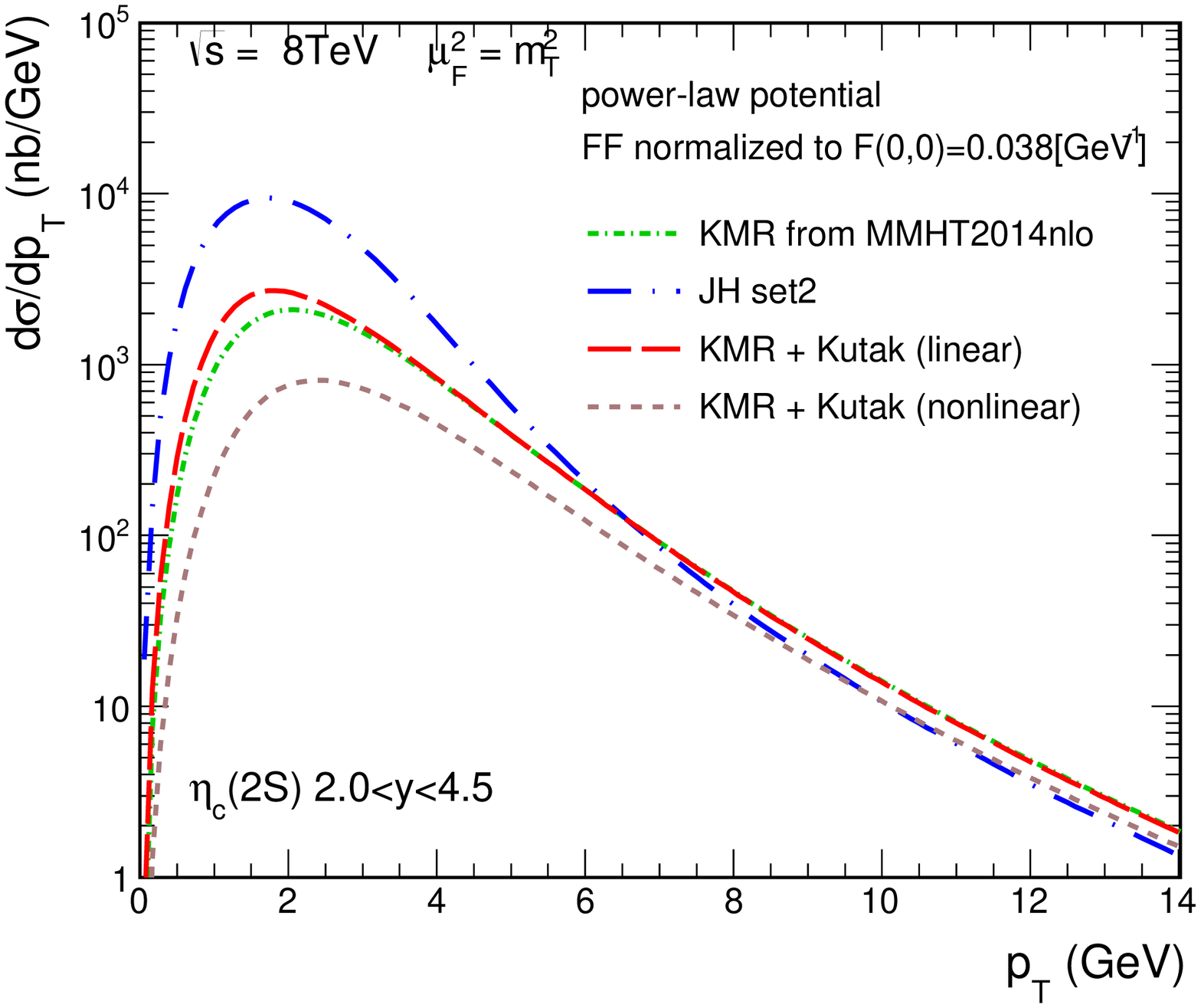}
    \includegraphics[width=0.32\textwidth]{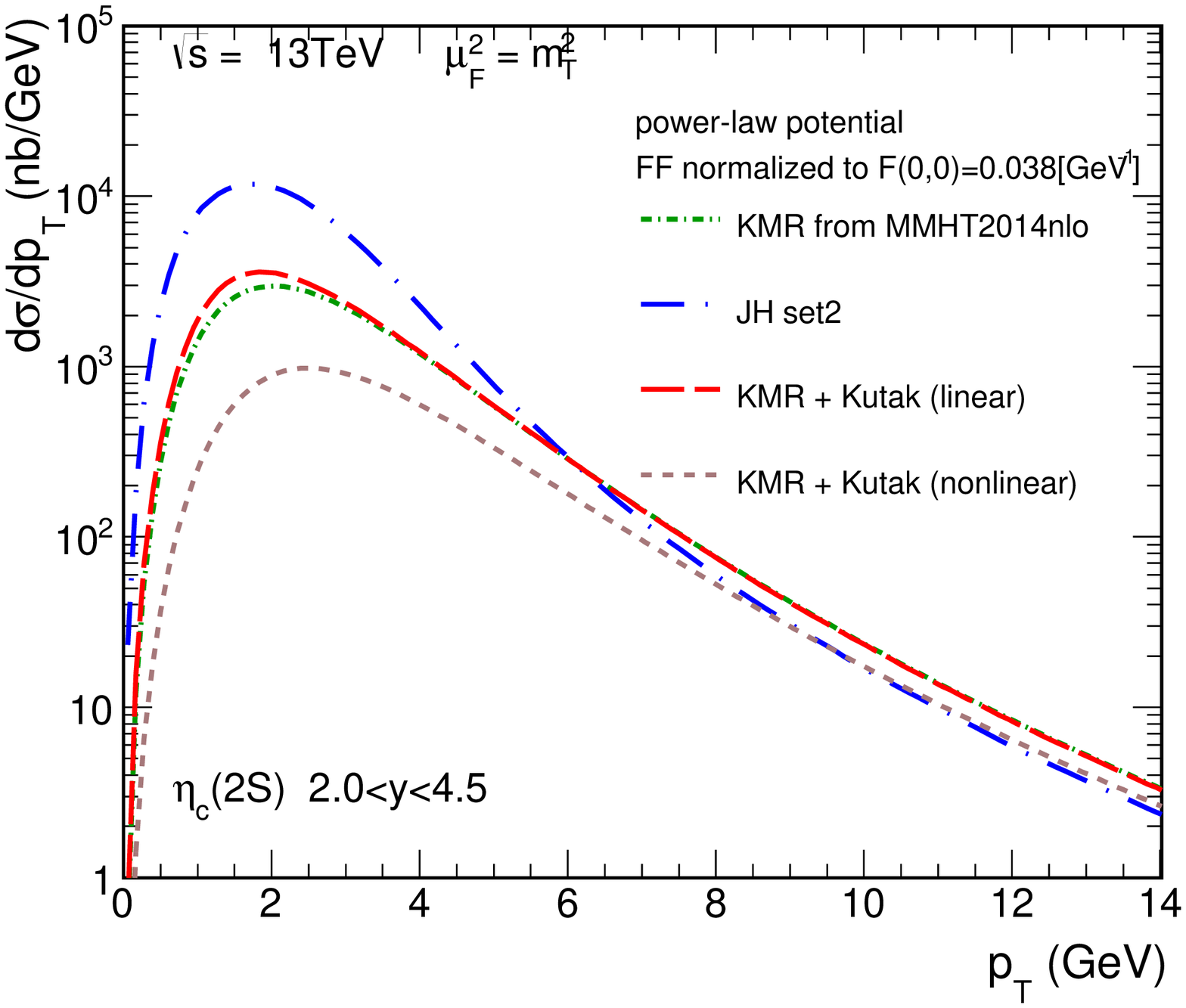}
    \caption{Differential cross section as a function of transverse
      momentum for prompt production of $\eta_{c}(2S)$
      for $\sqrt{s} = 7, 8, 13 \, \rm{TeV}$.
}
    \label{fig:dsig_dpt_etac2S}
\end{figure}

Our predictions for $\eta_c(2S)$ are shown in
Fig.\ref{fig:dsig_dpt_etac2S}. The shapes of the distributions are
similar to those for $\eta_c(1S)$ while the cross section is slightly
smaller.

\begin{figure}[h!]
    \centering
    \includegraphics[width=0.45\textwidth]{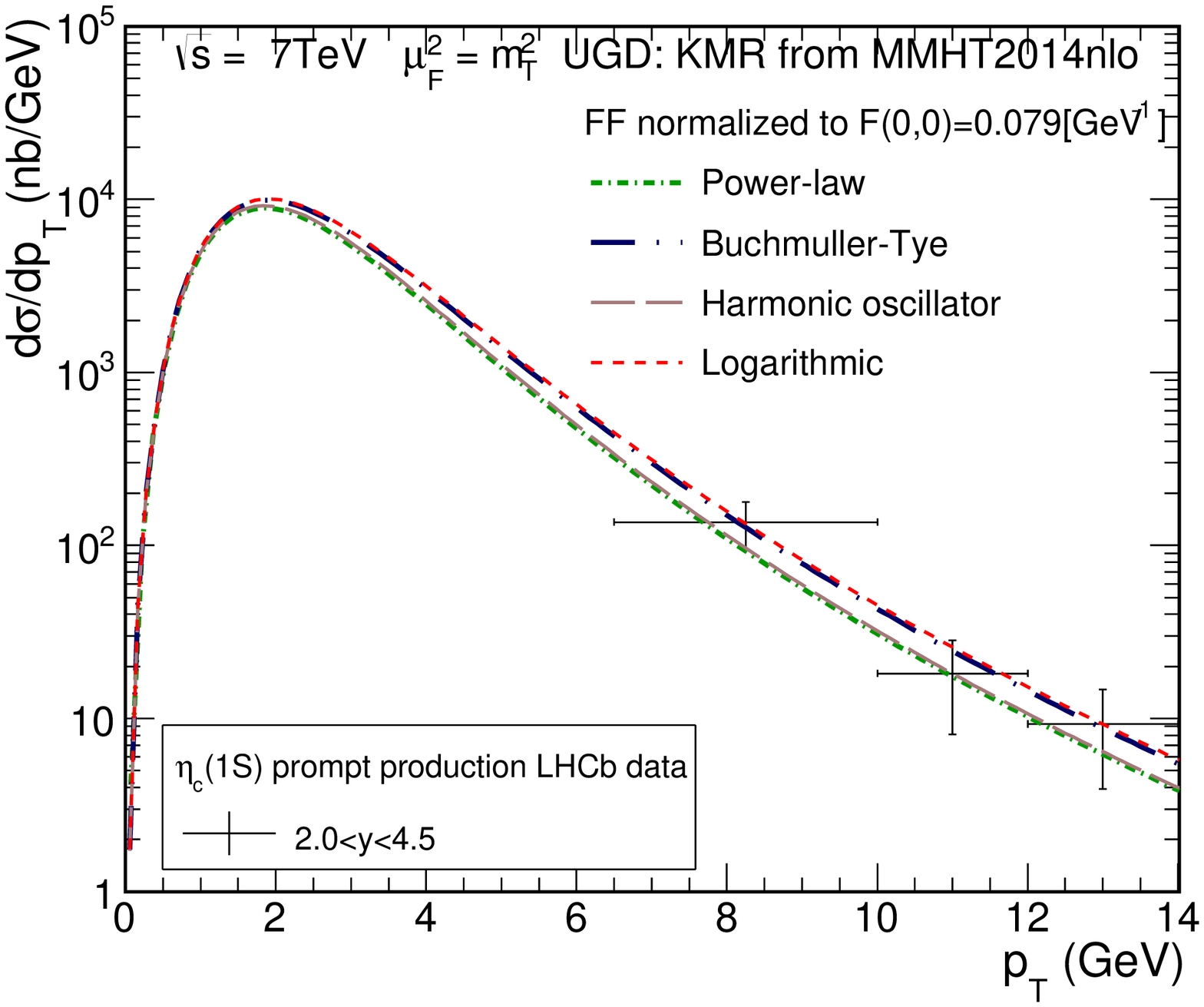}
    \includegraphics[width=0.45\textwidth]{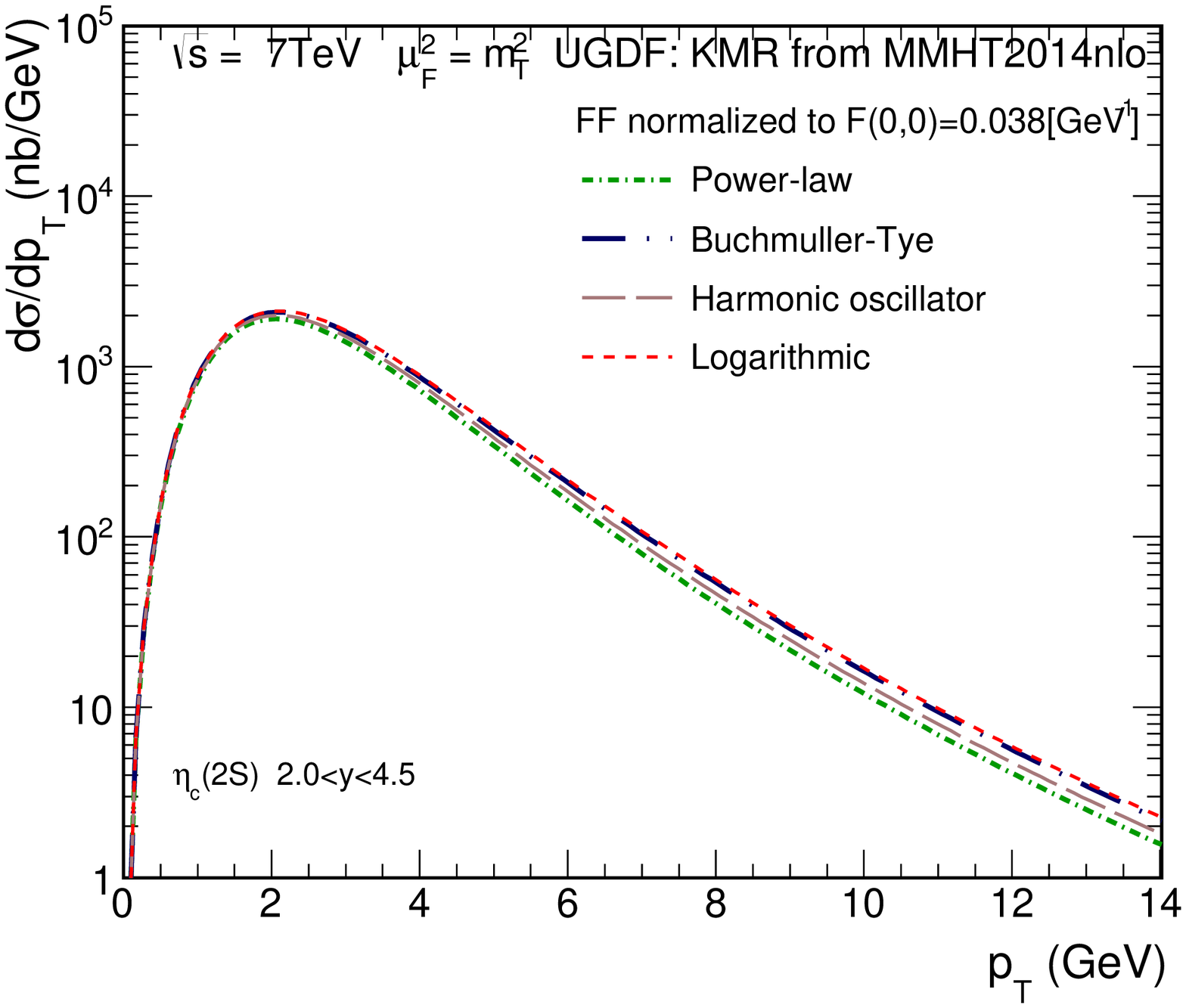}
    \caption{Transverse momentum distributions calculated with 
     different form factors obtained from different 
     potential models of quarkonium wave function and
     one common normalization of $|F(0,0)|$.}
    \label{fig:dsig_dpt_ff_exp}
\end{figure}

The dependence on form factors are shown in
Fig.\ref{fig:dsig_dpt_ff_exp}. In general our results are less uncertain
as far as the form factor is considered compared to uncertaintis due to
unintegrated gluon distributions shown above.

\begin{figure}[h!]
    \centering
    \includegraphics[width=0.45\textwidth]{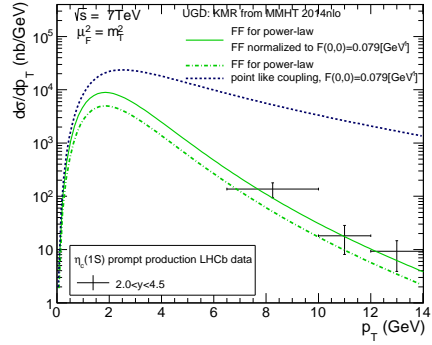}
    \caption{Comparison of results for two different transition form
      factor, computed with the KMR unintegrated gluon distribution.
      We also show result when the $(q_{1T}^2,q_{2T}^2)$ dependence
      of the transition form factor is neglected (short dashed line).}
    \label{fig:different_coupling}
\end{figure}

Finally in Fig.\ref{fig:different_coupling} we demonstrate how
important is inclusion of form factor. The effect is huge.
This puts into question all calculations in which the form factor is not
included.

\section{Conclusion}

Here we briefly summarize our results found in \cite{BGPSS2019}
and \cite{BPSS2020}.

\begin{itemize}

\item The transition form factor for different
wave functions obtained as a solution of the Schr\"odinger equation for 
the $c \bar c$ system, for different phenomenological $c \bar{c}$ potentials 
from the literature, was calculated.

\item We studied the transition form factors
for $\gamma^* \gamma^* \to \eta_c$ (1S,2S) for two space-like virtual 
photons, which can be accessed experimentally in future measurements 
of the cross section for the $e^+ e^- \rightarrow e^+ e^- \eta_c$
process in the double-tag mode.

\item The transition form factor for only one
off-shell photon as a function of its virtuality was studied and compared
to the BaBar data for the $\eta_c(1S)$ case.

\item
Predictions for $\eta_c(2S)$ were presented. 

\item 
Dependence of the transition form factor on the virtuality was studied 
and delayed convergence of the form factor to
its asymptotic value $\frac{8}{3}f_{\eta_{c}}$ as predicted by 
the standard hard scattering formalism, was presented.


\item There is practically 
no dependence of transition form factor on the asymmetry parameter $\omega$, 
which could be verified experimentally at Belle 2.

\item 
The $k_T$-factorization approach with modern UGDs lead to
good description of the LHCb data for 
$p p \to \eta_c(1S) \to p {\bar p}$ for $\sqrt{s}$ = 7, 8 TeV 
and somewhat worse for $\sqrt{s}$ = 13 TeV.
There is some room for color octet.
Feed down contribution is small \cite{Baranov:2019joi}.

\item Range of $x_1, x_2$ and $q_{1T}, q_{2T}$ was discussed.
For the LHCb kinematics very small longitudinal momentum
fractions are probed. Transverse momenta are not too small.

\item We do not see an obvious sign of the onset of saturation.
LHCb cross section grows even faster than our result without saturation.
However the gluon transverse momenta are not small.

\item Predictions for hadroproduction of $\eta_c(2S)$ were also presented.

\item We also discussed uncertainties related 
to $g^* g^* \to \eta_c$ form factor. They are somewhat smaller 
than those related to UGDs. 

\end{itemize}

Recently we have preformed similar sudies also for scalar quarkonium 
$\chi_c(0)$ \cite{BPSS2020_chic} and light $f_0(980)$ meson \cite{LMS2020}.

\section*{Acknowledgements}
I would like to thank Iza Babiarz, Victor Goncalves, Roman Pasechnik 
and Wolfgang Schafer for collaboration on the topics presented in this
short review.
This study was partially supported by the Polish National
Science Center grant UMO-2018/31/B/ST2/03537 and by the Center for
Innovation and Transfer of Natural Sciences and Engineering Knowledge in Rzesz\'ow.


\end{document}